\documentclass[useAMS,usenatbib]{mnras}
\usepackage{graphicx}
\usepackage{textcomp}
\usepackage{amssymb}\usepackage{hyperref}
\usepackage{amsmath,alltt}           
                                                                                                          
\usepackage{multirow}
\usepackage{rotating}
\usepackage{lscape}
\usepackage{times}
\usepackage{hyperref}
\usepackage{pdflscape}
\usepackage{supertabular}
\usepackage{verbatim}
\usepackage{appendix}
\usepackage{comment}

\newcommand\lsim{\mathrel{\rlap{\lower4pt\hbox{\hskip1pt$\sim$}}
        \raise1pt\hbox{$<$}}}
\newcommand\gsim{\mathrel{\rlap{\lower4pt\hbox{\hskip1pt$\sim$}}
        \raise1pt\hbox{$>$}}}
\newcommand{\be}{\begin{equation}}
\newcommand{\ba}{\begin{eqnarray}}
\newcommand{\ee}{\end{equation}}
\newcommand{\ea}{\end{eqnarray}}

\title[Thermodynamics of stellar multiplicity]{The thermodynamics of stellar multiplicity: the dynamical evolution of binary star populations in dense stellar environments}

\author[N. W. C. Leigh, N. C. Stone, J. J. Webb, W. Lyra]{N. W. C. Leigh$^{1,2}$, N. C. Stone$^{3}$, J. J. Webb$^{4}$, W. Lyra$^{5}$\\ 
$^{1}$Departamento de Astronom\'a, Facultad de Ciencias F\'sicas y Matem\'aticas, Universidad de Concepci\'on, Concepci\'on, Chile \\
$^{2}$Department of Astrophysics, American Museum of Natural History, New York, NY 10024, USA\\
$^{3}$Racah Institute of Physics, The Hebrew University, Jerusalem, 91904, Israel \\
$^{4}$ Department of Astronomy and Astrophysics, University of Toronto, 50 St. George Street, Toronto, ON, M5S 3H4, Canada \\
$^{5}$ New Mexico State University, Department of Astronomy, PO Box 30001, MSC 4500 Las Cruces, NM 88003, USA}

\begin{document}

\date{Accepted. Received; in original form}

\pagerange{\pageref{firstpage}--\pageref{lastpage}} \pubyear{2008}

\maketitle

\label{firstpage}

\begin{abstract}
We recently derived, using the density-of-states approximation, analytic distribution functions for the outcomes of direct single-binary scatterings \citep{stoneleigh19}. Using these outcome distribution functions, we present in this paper a self-consistent statistical mechanics-based analytic model obtained using the Fokker-Planck limit of 
the Boltzmann equation.  Our model quantifies the dominant gravitational physics, combining both strong and weak 
single-binary interactions, that drives the time evolution of binary orbital parameter distributions in dense stellar environments.  We focus in particular the distributions of binary orbital energies and eccentricities.  We find a novel steady state distribution of binary eccentricities, featuring strong depletions of both the highest and the lowest eccentricity binaries.  In energy space, we compare the predictions of our analytic model to the results of numerical N-body simulations, and find that the agreement is good for the initial conditions considered here.  This work is a first step toward the development of a fully self-consistent semi-analytic model for dynamically evolving binary star populations in dense stellar environments due to direct few-body interactions.
\end{abstract}


\section{Introduction}
\label{sec:intro}

Moore's Law is dead; the future will bring improved efficiency and speed to the computational sciences, but at a slower rate than before \citep[e.g.][]{wang15,wang16,bonetti20}.  Consequently, the demand for alternative models independent of computational limitations (e.g., analytic calculations) are becoming increasingly urgent for the first time in over half a century.  

Galactic and extragalactic globular clusters (GCs) are an example of an astrophysical system predominantly modeled today via complex numerical simulations.  One particular challenge in studying GCs is related to their collisional nature; central stellar densities are so high ($>$ 10$^5$ M$_{\odot}$ pc$^{-3}$) that direct encounters between single and binary stars occur frequently.  The exact rate depends on the host cluster properties but, for the densest GCs, the time-scale for direct single-binary and binary-binary encounters to occur is of order 1-10 Myr \citep[e.g.][]{leigh11,geller15}.  These two timescales are roughly equal for binary fractions $f_b$ $\sim$ 10\% \citep{sigurdsson93,leigh11}, such that the rate of single-binary interactions dominates over that of binary-binary interactions in clusters with $f_b \lesssim 10\%$.  These multibody interactions are expensive to resolve in direct N-body simulations, but play a key role in (i) determining the overall cluster evolution, due to the release of gravitational potential energy via ``binary  burning'' (as first suggested by \citealt{henon61}), and (ii) producing exotic sources of electromagnetic \citep{leigh11b, ivanova06, ivanova08} or gravitational wave \citep{portegies00, rodriguez16a, askar17, samsing18} radiation.

At present, a number of highly sophisticated computational tools have been developed to simulate the time evolution of dense stellar systems, including direct interactions involving binary stars.  Among the most successful of these are $N$-body simulations \citep[e.g.][]{aarseth75,aarseth85,aarseth99,aarseth03}, which calculate directly the gravitational acceleration exerted on every particle in the system, summed over all other particles.  The code then propagates the system forward through time using an appropriately chosen time-step, and step-by-step the cluster is evolved.  $N$-body simulations originated over half a century ago \citep[e.g.][]{vonhoerner60,vonhoerner63,aarseth63}.  They have been evolving ever since, growing increasingly sophisticated over time.  Many new software and hardware techniques have been introduced and incorporated, including three-body regularization \citep{aarseth74}, chain regularization \citep{mikkola93,mikkola98}, the Hermite scheme \citep{makino92}, the GRAPE hardware \citep{sugimoto90,makino96,nitadori12} and, more recently, the introduction of GPUs for accelerated computing.  Due to the increased computational expense, $N$-body simulations struggle to simulate self-gravitating systems with large particle numbers (this includes the largest globular clusters, but also most nuclear star clusters) and high binary fractions.  For this general category of initial conditions (i.e., massive clusters with high binary fractions), long simulation run times, of order a year, can be needed to complete even a single simulation \citep[e.g.][]{wang15,wang16}.

Monte Carlo (MC) simulations for GC evolution have also proven highly successful, covering a range of parameter space inaccessible to more computationally-expensive $N$-body simulations.  MC models rely on statistical approximations to calculate the time-dependent diffusion of energy throughout cluster due to two-body relaxation.  Consequently, MC models are able to handle much larger particle numbers, including larger numbers of binaries, and hence are able to simulate more massive and denser GCs than are $N$-body models, at a significantly reduced computational expense.  MC models are incredibly fast; they are able to evolve a million cluster stars for a Hubble time in a matter of days (roughly two orders of magnitude faster than state-of-the-art N-body simulations).  The MC method goes back to the pioneering works of \citet{henon71} and \citet{spitzer71}, with many other researchers later building upon these earliest ideas \citep[e.g.][]{shapiro78,stodolkiewicz82,stodolkiewicz86,giersz98,giersz01}.  Modern MC codes supplement their diffusive treatment of bulk cluster evolution with embedded few-body integrators, most notably FEWBODY \citep{fregeau04}. These are typically used to treat direct three- (i.e., single-binary) and four-body (i.e., binary-binary) interactions in MC simulations \citep{giersz08,downing10,hypki13,leigh15}.  As a result, MC models are  useful for simulations of massive star clusters, capturing the time evolution of temporarily bound and highly chaotic three- and four-body systems.  These are thought to be the main source of stellar collisions and/or mergers in dense star clusters and an important formation channel for many exotic stellar systems \citep{leonard89,pooley06,sills97,sills01,leigh11,leigh12,leigh18b}.    

In spite of these successes, MC models are still limited, in that they cannot simulate small particle numbers (i.e., $N \lesssim 10^4$) \citep[e.g.][]{giersz98,giersz01,kremer20}.  Thus, we are still lacking a single tool capable of covering the entire range of parameter space relevant to real star clusters, ranging from open to globular and even nuclear clusters. 
While $N$-body and MC simulations often find good agreement for bulk cluster properties \citep{giersz13}, detailed comparisons find areas of disagreement, such as phases of deep core collapse \citep{rodriguez16b}.  The assumptions of existing MC codes also prevent simulation of some clusters of interest, for example those with net rotation\footnote{Although see \citet{fiestas06, kim08} for an example of an MC code which can simulate rotating clusters, albeit at the cost of a special type of assumed isotropy.} (which can accelerate core collapse, possibly increasing the rates of exotica production, \citealt{ernst07}) or those where vector resonant relaxation has created strong deviations from spherical symmetry \citep{meiron19, szolgyen19}.  Such clusters nonetheless contain collisionally evolving binaries, and it would be useful to have a computationally efficient tool with which to model them.  

Another limitation of MC codes is that three- and four-body interactions, along with stellar and binary evolution, are assumed to occur in isolation, and so do not allow for the interruption of ongoing interactions and other perturbative effects that occur in a live star cluster environment \citep[e.g.][]{geller15,leigh16c}.  For example, binaries are continually perturbed by the distant flybys of other stars in the cluster.  These perturbations happen at every time-step in an $N$-body code, often leading to random walks in the orbital eccentricity that drive mergers.  MC codes do not capture this aspect of the N-body dynamics \citep{giersz98,giersz08,kremer20}, again suggesting the need for an additional efficient computational tool.

Beyond the successes and limitations
of the aforementioned computational tools for simulating GC evolution, we are still lacking a robust and 
transparent
\textit{physical} model that can be used to understand the dominant physical processes dynamically re-shaping the properties of binary populations over cosmic time.  In other words, we can give a set of initial conditions to a computer simulation and it will compute for us the final result.  But how do we understand, and even quantitatively characterize, what this output is telling us about everything that happened in between providing the input and reading the output?  

In this paper, we directly address these practical and conceptual issues, by formulating a self-consistent statistical mechanics model 
to describe the time evolution of the properties of a population of binaries, based on a master-type equation.  Our semi-analytic model quantifies the dominant gravitational physics driving the time evolution of the binary orbital parameter distributions in dense stellar environments, namely direct three-body interactions between single and binary stars.  A practical solution of our master equation can be obtained in the Fokker-Planck limit, which we 
complete using the 
formulation for the outcomes of chaotic three-body interactions introduced in \citet{stoneleigh19}.  
We compare the results of our analytic calculations to a suite of N-body simulations, and show that the agreement is excellent for the range of initial conditions considered here.
In an Appendix, we go on to discuss how to adapt our base model to also include four-body or binary-binary interactions.  Although four-body interactions are not always dominant over three-body interactions, they are always occurring for non-zero binary fractions, and contribute to the underlying dynamical evolution of the binary orbital parameter distributions.  Hence, a complete model must include this contribution which, as described in the Appendix, will be the focus of future work.  In principle, this will allow us to address the key thermodynamic issue of whether or not (and on what timescale) stellar multiplicity in dense cluster environments ever reaches a steady- or equilibrium-state is discussed, as quantified by the ratio of single, binary and triple stars.  In steady-state, their relative fractions should remain approximately constant.

\section{The Model} \label{model}

In this section, we present our model for dynamically evolving in time the distribution functions for an entire population of binary star systems in dense stellar environments, accounting for the effects of repeated binary-single encounters.  In other words, we model the time evolution of a population of binaries embedded in a ``heat bath'' of single stars.  This formalism is similar in spirit to the pioneering early work of \citet{goodman93}, but it is both (i) more general, in that we present a multi-dimensional version of their underlying master equation, and (ii) also more accurate, in that we take advantage of recent advances in the statistical theory of chaotic multi-body encounters (\citealt{stoneleigh19}, but see also \citealt{ginat20, kol20, manwadkar21}) and the secular theory of weak encounters \citep{hamers19a, hamers19b}.  We begin with a simple, 1D, second-order (Fokker-Planck) method to quantify the time evolution of the binary orbital energy distribution. We then present the more complete multi-dimensional master equation that describes the full time evolution of all binary variables of interest.  
This equation is complex enough that we do not solve it fully in this paper, but we show how it can reduce to the simplified Fokker-Planck limit presented earlier, and analyze various 1D solutions.
We also present a simple model to quantify the backreaction effects on the properties of the host star cluster (e.g., the time evolution of the core radius, the time at which core-collapse occurs, and so on).

\subsection{A Fokker-Planck equation in energy space} \label{boltzmann}

We begin by reviewing the features of a 1D Fokker-Planck equation valid to second-order (in the fractional changes of the variable of interest).  In the next subsections, we show that this formalism is the relevant limit for some aspects of a more general master equation for binary evolution.  Since this is the primary limit of binary evolution we will investigate in this paper, we begin with a brief review, focusing on distributions of binary energy $E_{\rm B}$ for specificity.

Assuming diffusive time evolution of the binary orbital energy probability distribution function $\mathcal{N}(E_{\rm B}) = {\rm d} N_{\rm B} / {\rm d}E_{\rm B}$, we can apply a standard Fokker-Planck equation:
\begin{align}
\frac{\partial \mathcal{N}}{\partial t} = & -\frac{\partial }{\partial E_{\rm B}}\Big[ \mathcal{N}(E_{\rm B})\langle {\Delta}E_{\rm B}\rangle \Big] \label{eq:FP1} \\ 
&+ \frac{1}{2}\frac{\partial ^2}{\partial E_{\rm B}^2}\Big[\mathcal{N}(E_{\rm B})\langle {\Delta}E_{\rm B}^2\rangle \Big], \notag
\end{align}
where $\langle {\Delta}E_{\rm B} \rangle$ and $\langle {\Delta}E_{\rm B}^2 \rangle$ are the first- and second-order diffusion coefficients.  The boundary conditions of the equation above are set by the cluster hard-soft boundary (i.e., $|E_{\rm B}| = |E_{\rm HS}| = \frac{1}{2}\bar{M}\sigma^2$, where $\bar{M}$ is the mean stellar mass in the cluster) at the loosely bound end of the distribution, and by the criterion for a contact binary (i.e., $|E_{\rm B}| = |E_{\rm coll}| \approx  \frac{GM^2}{R}$, where $M$ is the mass and $R$ is the radius of the test star species) at the compact end.

This equation arises from the assumption that the binary orbital energy distribution function evolves diffusively in its host cluster.  In this way, changes to this distribution can be described as a local process, with binaries flowing mostly in to and out of adjacent energy bins.  
We can understand the origin of the Fokker-Planck equation as follows.  If we consider a general process of binary evolution through energy space, then the binary energy distribution at a time $t+\Delta t$ will be
\begin{equation}
\label{eqn:eqboltzmann2}
\mathcal{N}(E_{\rm B},t+{\Delta}t) = \int \mathcal{N}(E_{\rm B}-{\Delta}E_{\rm B},t)P(E_{\rm B}-{\Delta}E_{\rm B},{\Delta}E_{\rm B})d{\Delta}E_{\rm B},
\end{equation}
where the transfer function,
\begin{equation}
\label{eqn:transferfunc}
P(E_{\rm B}-{\Delta}E_{\rm B},{\Delta}E_{\rm B}) = \Gamma(E_{\rm B},E_{\rm 0})\mathcal{F}(E_{\rm B},E_{\rm 0}) \Delta t,
\end{equation}
represents the probability that in a differential time interval $\Delta t$, a binary will reach a final energy $E_{\rm B}$ from an initial energy $E_{\rm B} - \Delta E_{\rm B}$.
Here the function $\mathcal{F}(E_{\rm B},E_{\rm 0})$ is the differential probability of a single ``encounter'' transitioning a test  binary from initial energy $E_0 = E_{\rm B}-\Delta E_{\rm B}$ to a final energy $E_{\rm B}$, and $\Gamma$ is the rate of all such encounters (here our language is general, but our eventual application will be to consider ``encounter'' arising from single-binary encounters).  
Following \citet{spitzer87} and Taylor expanding Equation~\ref{eqn:eqboltzmann2}, we obtain:
\begin{align}
&\mathcal{N}(E_{\rm B},t) + {\Delta}t\frac{\partial \mathcal{N}}{\partial t} = \\ &\int \Big[\mathcal{N}(E_{\rm B})P(E_{\rm B}-{\Delta}E_{\rm B},{\Delta}E_{\rm B})\notag \\
 & - \frac{\partial }{\partial E_{\rm B}}\Big(\mathcal{N}(E_{\rm B},t)P(E_{\rm B}-{\Delta}E_{\rm B},{\Delta}E_{\rm B})\Big){\Delta}E_{\rm B}\notag \\
 & + \frac{1}{2}\frac{\partial ^2}{\partial E_{\rm B}^2}\Big(\mathcal{N}(E_{\rm B},t)P(E_{\rm B}-{\Delta}E_{\rm B},{\Delta}E_{\rm B})\Big){\Delta}E_{\rm B}^2\Big]d{\Delta}E_{\rm B}, \notag
\end{align}
which can, after eliminating a single factor of $\mathcal{N}(E_{\rm B}, t)$ from both sides, be simplified to Eq. \ref{eq:FP1}.
Here we now see the origin of the diffusion coefficients in our assumption that the relevant ``encounters'' are perturbative in nature, or in other words that our Taylor expansion was validly truncated at second order:
\begin{equation}
\label{eqn:diffusion1}
\langle{\Delta}E_{\rm B}\rangle = \int \Gamma(E_{\rm B},E_{\rm 0})\mathcal{F}(E_{\rm B},E_{\rm 0}){\Delta}E_{\rm B}d{\Delta}E_{\rm B},
\end{equation}
and
\begin{equation}
\label{eqn:diffusion2}
\langle{\Delta}E_{\rm B}^2\rangle = \int \Gamma(E_{\rm B},E_{\rm 0})\mathcal{F}(E_{\rm B},E_{\rm 0}){\Delta}E_{\rm B}^2d{\Delta}E_{\rm B}.
\end{equation}
Equation~\ref{eq:FP1} is a partial differential equation that can be solved numerically.  To do this, we must close the system by writing the per-binary encounter rate.  Focusing on strong single-binary scatterings for specificity, the gravitationally focused scattering rate is:
\begin{equation}
\label{eqn:mfp2}
\Gamma = \frac{3{\pi}n_{\rm s}G^2(m_{\rm B}+m_{\rm s})m_{\rm B}}{{\sigma}|E_{\rm B}-{\Delta}E_{\rm B}|},
\end{equation}
where $\sigma$ is the cluster velocity dispersion, $m_{\rm B}$ and $m_{\rm s}$ are the binary and single star masses, respectively, and $n_{\rm s}$ is the density of scatterers. 
The total encounter energy can then be written:
\begin{equation}
\label{eqn:entot}
E_{\rm 0} = E_{\rm B} - {\Delta}E_{\rm B} + \frac{1}{2}\frac{m_{\rm B}m_{\rm s}}{m_{\rm B}+m_{\rm s}}\sigma^2
\end{equation}

The distribution of binary orbital energies for hard binaries left over after single-binary interactions is often approximated as a power law: \citep{monaghan76a,monaghan76b,valtonen06}:
\begin{equation}
\label{eqn:outcome}
\mathcal{F}_{\rm VK}(|E_{\rm B}|)d|E_{\rm B}| = (n-1)|E_{\rm 0}|^{n-1}|E_{\rm B}|^{-n}d|E_{\rm B}|,
\end{equation}
The parameter $n$ depends on the total interaction angular momentum $L$, and has been fit to numerical simulations as \citep{valtonen06}:
\begin{equation}
\label{eqn:n}
n = 3 + 18\tilde{L}^2,
\end{equation}
where $\tilde{L}$ is a normalized version of the total encounter angular momentum (see \citet{valtonen06} for more details).  
Alternatively, one may use first-principles estimates for $\mathcal{F}$ from the ergodic formalism of \citet{stoneleigh19}, which will be our approach later in this paper.  With $\mathcal{F}$ specified,
Equations~\ref{eqn:mfp2} and~\ref{eqn:entot} can be plugged in to Equation~\ref{eq:FP1} such that it becomes straight-forward to solve numerically.

\subsection{The Master Equation and its Fokker-Planck Limit} \label{master}

In this section, we further generalize our formulation to include angular momentum, moving away from the quasi-empirical outcome distribution provided in \citet{valtonen06} and toward the more robust formulation provided in \citet{stoneleigh19}.  The latter self-consistently accounts for angular momentum in the density-of-states formulation, whereas the former uses numerical scattering experiments to obtain approximate outcome distributions as a function of the total angular momentum.  We will also present the evolution of the binary distribution from a first-principles kinetic perspective, by working in the Fokker-Planck limit of a master equation.

We consider the evolution of a population of binaries inside a dense, collisional, spherically symmetric star cluster.  The cluster has an ambient density of single stars $n(r)$ and a velocity dispersion $\sigma(r)$ that are both assumed to vary with radius $r$.  We assume for now that the cluster is relaxed and isotropic, so that the stellar velocity distribution is Maxwellian,
\begin{equation}
    f(v) = \sqrt{ \frac{2}{\pi} } \frac{v^2}{\sigma^3} \exp(-v^2 / 2\sigma^2). \label{eq:maxwellian}
\end{equation}
The binaries in this cluster can be characterized by six binary orbital elements, $\vec{B}$.  For practical purposes, we focus on their internal energy $E_{\rm B}$, internal angular momentum $L_{\rm B}$ (or equivalently their semimajor axis $a_{\rm B}$ and eccentricity $e_{\rm B}$), and orbital orientation $C_{\rm B}$; other, angular orbital elements can generally be assumed to be distributed isotropically.  We define $C_{\rm B} = \cos I_{\rm B}$, where $I_{\rm B}$ is the inclination angle between the binary angular momentum vector and an arbitrary reference direction.  

The number of binaries with energy $E_{\rm B}$, angular momentum $L_{\rm B}$, and orientation $C_{\rm B}$ within an infinitesimal range ${\rm d}E_{\rm B}{\rm d}L_{\rm B}{\rm d}C_{\rm B}$ is  $\mathcal{N}(E_{\rm B}, L_{\rm B}, C_{\rm B}){\rm d}E_{\rm B}{\rm d}L_{\rm B}{\rm d}C_{\rm B}$.  The binary probability distribution\footnote{Note that throughout this paper, we use the symbol $\mathcal{N}$ to denote distributions of binary orbital elements over varied numbers of dimensions.  However, the dimensionality of the distribution should always be clear from either context or explicit labeling of its arguments.} $\mathcal{N}(E_{\rm B}, L_{\rm B}, C_{\rm B})$ will evolve due to scatterings with single stars and with each other.  These scatterings may be weak, distant encounters, in which case the tools of secular theory can be used to understand the exchange of energy and angular momentum \citep{HeggieRasio96}.  Alternatively, some scatterings may form temporarily bound (``resonant'') triple systems, which will eventually disintegrate into a survivor binary and an escaping single star; the outcomes of these strong scatterings can be understood through the ergodic hypothesis \citep{stoneleigh19}.

The long-term evolution of a single-mass binary distribution function will be governed by a master equation:
\begin{align}
    \frac{\partial \mathcal{N}}{\partial t} = & \idotsint {\rm d}^6{\Delta \vec{B}}\Big[\Psi(\vec{B}-\Delta \vec{B}, \Delta\vec{B})\mathcal{N}(\vec{B}-\Delta\vec{B})\\
    & - \Psi(\vec{B}, \Delta\vec{B})\mathcal{N}(\vec{B}) \Big] - \left(\frac{\partial \mathcal{N}}{\partial t} \right)_{\rm sink}. \label{eq:masterEq}
\end{align}
Here $\Psi(\vec{B}, \Delta\vec{B})$ is a transition probability describing the differential rate that stars are scattered into the phase space region $\vec{B}+\Delta\vec{B}$ from the original phase space coordinates $\vec{B}$, and the final term is a catchall ``sink'' representing ways binaries may be destroyed.  Equation \ref{eq:masterEq} is the most general kinetic formulation for the local evolution of a population of single-mass binaries, but if we assume isotropy, we need only to consider a two-dimensional integral and a two-dimensional distribution $\mathcal{N}(E_{\rm B}, L_{\rm B})$.  

In principle, one may Taylor-expand Equation \ref{eq:masterEq} in the small $\Delta\vec{B}$ limit, and obtain a Fokker-Planck equation generalizing Eq. \ref{eq:FP1}:
\begin{align}
    \frac{\partial \mathcal{N}}{\partial t} = & -\frac{\partial}{\partial E_{\rm B}}(\langle \Delta E_{\rm B} \rangle \mathcal{N}) + \frac{1}{2}\frac{\partial^2}{\partial^2 E_{\rm B}}(\langle (\Delta E_{\rm B})^2 \rangle \mathcal{N})\notag  \\
   & -\frac{\partial}{\partial L}(\langle \Delta L_{\rm B} \rangle \mathcal{N})  + \frac{1}{2}\frac{\partial^2}{\partial^2 L_{\rm B}}(\langle (\Delta L_{\rm B})^2 \rangle \mathcal{N}) \label{eq:FP} \\
    &  + \frac{\partial^2}{\partial E_{\rm B}\partial L_{\rm B}}(\langle \Delta E_{\rm B}\Delta L_{\rm B} \rangle \mathcal{N}) - \left(\frac{\partial \mathcal{N}}{\partial t} \right)_{\rm sink}. \notag
\end{align}
Here terms such as $\langle \Delta E \rangle$ are effective diffusion coefficients.  These diffusion coefficients (and the transition probability $\Psi$ from which they originate) represent the cumulative effect of interactions between a population of binaries and their perturbers.  This includes both self-interactions (i.e. binary-binary scatterings) and external interactions (e.g. binary-single scatterings).  In this work, we will focus only on binary-single scatterings, which dominate the rate of binary evolution so long as the binary fraction is sufficiently low $\lesssim$ 10\% \citep[e.g.][]{sigurdsson93,leigh11}.

Qualitatively different types of binary-single scatterings are possible, which we can break down into four categories: (i) ionizations, (ii) flybys, (iii) prompt exchanges, and (iv) resonant encounters.  A closer look at some of these categories reveals the inconsistency of the (full) Fokker-Planck limit of the master equation: while energy ($E_{\rm B}$) evolution is indeed diffusive, angular momentum ($L_{\rm B}$) evolves in a strongly non-diffusive way during resonances and prompt exchanges.  We will postpone a full solution of the master equation for future work.  In this paper, we work with a hybrid, ``two-timescale'' approach to understand the evolution of $E_{\rm B}$ and $L_{\rm B}$ separately.  For simplicity, we will treat ionizations by assuming that all binaries with energy below the hard-soft boundary are promptly ionized.

\subsection{Binary energies}

Here we will neglect flybys and prompt exchanges, since we are concerned with hard binaries (soft binaries are short-lived and quickly ionized).  In hard binaries, resonant and non-resonant encounters contribute roughly equally to total energy evolution \citep{hut84}, although resonant encounters do dominate the largest energy shifts.  We will thus only consider energy evolution through resonant encounters, an approximation that should be valid at a factor $\approx 2$ level, and which is motivated primarily by the lack of an analytic formalism for energy exchange in strong but non-resonant encounters \citep{heggie93} \footnote{Such a formalism exists for weak non-resonant encounters \citep{heggie75, heggie93}, but these are always sub-dominant in energy evolution.}.  For simplicity, we treat ionizations by assuming that all binaries with energy below the hard-soft boundary are promptly ionized.

The mean change in binary energy in a given resonant encounter, with conserved energy $E_0$, conserved angular momentum $L_0$, and a mass triplet $\vec{m} = \{m_1, m_2, m_3\}$ is
\begin{equation}
    \Delta E_{\rm B} = \iiint E_{\rm B} \mathcal{T}(E_0, L_0, \vec{m}){\rm d}E_{\rm B}{\rm d}L_{\rm B}{\rm d}C_{\rm B}. \label{eq:situationalDC}
\end{equation}
Here we have used the ergodic outcome distribution, $\mathcal{T}(E_0, L_0, \vec{m}) = {\rm d}V / {\rm d}E_{\rm B}{\rm d}L_{\rm B}{\rm d}C_{\rm B}$, which partitions outcomes of non-hierarchical triple disintegration uniformly across a high-dimensional phase volume ($V$), ultimately giving non-trivial outcome distributions in the survivor binary's $E_{\rm B}$, $L_{\rm B}$, and cosine-inclination $C_{\rm B}$ \citep{stoneleigh19}. 

While Equation \ref{eq:situationalDC} gives a moment of the outcome distribution for a particular combination of $E_0$, $L_0$, and $\vec{m}$ (and can easily be generalized to produce $\Delta L_{\rm B}$, $(\Delta E_{\rm B})^2$, etc.) we are interested in computing rate-averaged diffusion coefficients.  If resonant encounters happen at a differential rate ${\rm d}\Gamma / {\rm d}E_0{\rm d}L_0{\rm d}C_0$, the rate-averaged $k$th diffusion coefficient in energy is
\begin{equation}
    \langle \Delta E_{\rm B}^k \rangle = \iiint \Delta E_{\rm B}^k(E_0, L_0, \vec{m}) \frac{{\rm d}\Gamma}{{\rm d}E_0{\rm d}L_0{\rm d}C_0}{\rm d}E_0{\rm d}L_0{\rm d}C_0.
\end{equation}
Here we have introduced an additional variable absent from $\mathcal{T}$, $C_0$, which represents the cosine of the inclination between the pre-encounter binary's orbital plane, and the mutual orbital plane between the binary and the single star it is encountering.  

Differential encounter rates are more easily expressed in terms of the impact parameter $b$ and relative velocity at infinity, $v_\infty$.  Specifically, the differential encounter rate is:
\begin{equation}
\label{eqn:gammanick}
    \frac{{\rm d}\Gamma}{{\rm d}v_\infty {\rm d}b {\rm d}C_0} = 2\pi n b.
\end{equation}
Here $n$ is the local number density of single stars.  These variables are related to $E_0$ and $L_0$ through the following equations:
\begin{align}
    E_0 = & E_{\rm b} + \frac{1}{2}\mu v_\infty^2 \label{eq:energydef}\\
    L_0 = & (L_{\rm b}^2 + \mu^2b^2v_\infty^2 + 2\mu b v_\infty C_0 L_{\rm b})^{1/2},
\end{align}
where $\mu = m_{\rm b} m_3 / (m_{\rm b}+m_3)$ is the reduced mass of the encounter, and we have denoted the variables of the pre-scattering binary with lower-case ``b'' subscripts.  Performing a change of variables, we find that
\begin{equation}
    d\Gamma = \frac{\pi n L_0 f(E_0, E_{\rm b})}{(2\mu (E_0 - E_{\rm b}) )^{3/2}} \left( 1 -\frac{C_0 L_{\rm b} }{\sqrt{L_0^2 - L_{\rm b}^2(1-C_0^2)}} \right) {\rm d}E_0 {\rm d}L_0 {\rm d}C_0.
\end{equation}
Here we have assumed that $v_\infty$ is drawn from a velocity distribution $f(v)$ (which can be written as a function of $E_0$ and $E_{\rm b}$ through Eq. \ref{eq:energydef}), although we have left this general for the moment (rather than specifying a Maxwellian).  Integrating this differential encounter rate ${\rm d}C_{\rm 0}$ (from $-1$ to $1$, i.e. under the assumption of velocity isotropy) eliminates the $L_{\rm b}$ dependence:
\begin{equation}
\label{eqn:FP}
    {\rm d}^2\Gamma = \frac{2\pi n L_0 f(E_0, E_{\rm b})}{(2\mu (E_0 - E_{\rm b}) )^{3/2}}{\rm d}E_0 {\rm d}L_0.
\end{equation}
Eq. \ref{eq:FP} is, unfortunately, not valid across all timescales.  While binary energy evolution is a fundamentally diffusive process (even strongly resonant encounters rarely change individual binary energies by more than a factor $\approx 2$; \citealt{hut84, stoneleigh19}), resonances lead to highly non-diffusive evolution of binary angular momentum\footnote{Unless the binary is much higher mass than the population of field stars it scatters against; however, this more extreme mass ratio limit greatly reduces the importance of resonant scatterings.}.  We therefore analyze two different limits of Eq. \ref{eq:FP}: first, a relatively simple, 1D equation in energy space, where we assume a steady-state eccentricity distribution.  In this limit, we have
\begin{equation}
    \frac{\partial \mathcal{N}}{\partial t} =  -\frac{\partial}{\partial E_{\rm B}}(\langle \Delta E_{\rm B} \rangle \mathcal{N}) + \frac{1}{2}\frac{\partial^2}{\partial^2 E_{\rm B}}(\langle \Delta E_{\rm B}^2 \rangle \mathcal{N}) - \left(\frac{\partial \mathcal{N}}{\partial t} \right)_{\rm sink}.  \label{eq:FP2}
\end{equation}
We allow $E_{\rm B}$ to range from $E_{\rm HS}$ to $E_{\rm coll}$.  
At these boundaries, we impose a Dirichlet-type boundary condition, with $N(E_{\rm B})=0$.  Alternatively, one could allow soft binaries to be included, in which case the limits would become 0 and $E_{\rm coll}$.  We neglect soft binaries because of their rapid rate of ionization, but this could in the future be modeled with an appropriate choice of volumetric sink function.

\begin{figure}
\begin{center}
\includegraphics[width=\columnwidth]{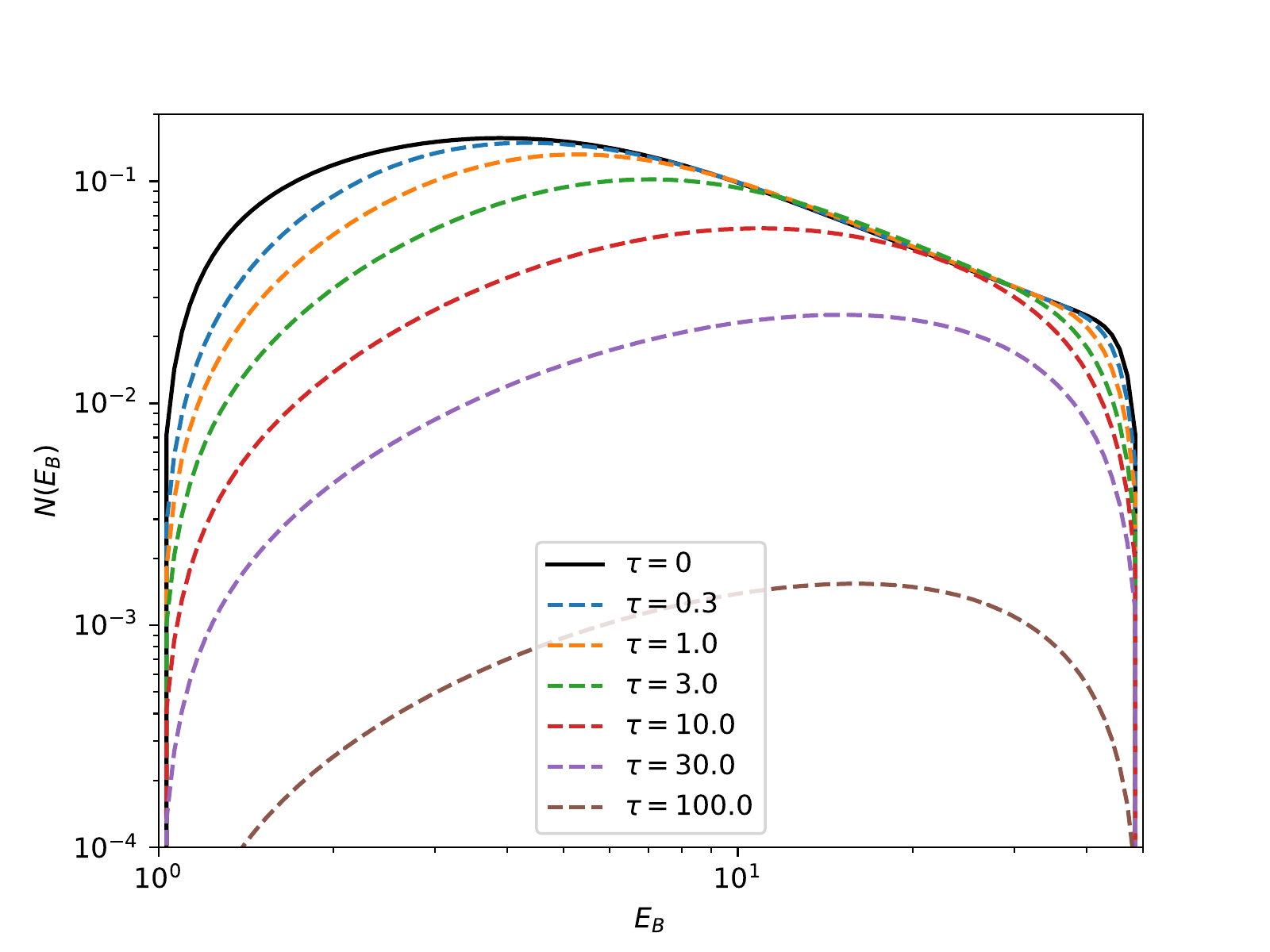}
\end{center}
\caption{Numerical results for the solution of Equation \ref{eq:FP2}.  Different colors show snapshots of the 1D binary distribution function $\mathcal{N}$($E_{\rm B}$) at different dimensionless times $\tau$, which is physical time normalized by units of the local (core) relaxation time.  Initial conditions (taken from \"Opik's Law, modified by exponential cutoffs near boundaries) are shown as a solid black line, while later steps in the evolution of the binary distribution function have been shown as colored dashed lines.  The absorbing boundary conditions are dimensionless (i.e., normalized by the hard-soft boundary energy $E_{\rm HS}$) where the binary energies $E_{\rm B}=1$ and $E_{\rm B}=50$ represent the hard-soft boundary and the point of direct binary collisions, respectively.}
\label{fig:FokkerPlanckSolution}
\end{figure}

Equation~\ref{eq:FP2} is a partial differential equation that can be solved numerically.  To do this, it is necessary to write explicit diffusion coefficients.  We find remarkably simple rates of diffusion across the space of orbital elements using $\mathcal{T}$ taken from \citet{stoneleigh19}, assuming a Maxwellian distribution of relative velocities, and considering particles all of equal mass, we find that 
\begin{equation}
    \langle \Delta E_{\rm B}\rangle = \frac{3}{64} \sqrt{\frac{\pi}{2}} \frac{G^2 n m^3}{\sigma |E_{\rm B}|} \Big(E_{\rm B} + m\sigma^2(1-\exp(E_{\rm B}/m\sigma^2)) \Big)
\end{equation}
and
\begin{align}
        \langle \Delta E_{\rm B}^2\rangle =&  \frac{1}{32}\sqrt{\frac{\pi}{2}} \frac{G^2 n m^3}{\sigma |E_{\rm B}|} \Big(2m\sigma^2 E_{\rm B} \notag \\ & +E_{\rm B}^2 + 2m^2\sigma^4(1-\exp(E_{\rm B}/m\sigma^2)) \Big).
\end{align}
The one approximation needed to derive these diffusion coefficients is to approximate $\iint \mathcal{T} {\rm d}L_{\rm B} {\rm d}C_{\rm B} = {\rm d}V/{\rm d}E_{\rm B} \propto E_{\rm B}^{-4}$, an approximate scaling that is quite accurate for equal-mass resonant scatterings \citep{stoneleigh19}.
Numerical results are shown in Fig. \ref{fig:FokkerPlanckSolution}.  We see that the binary population steadily depletes over time, initially at low energies but eventually at higher energies as well.  Eventually, a quasi-steady state energy distribution is reached, with a constant shape but continuously decaying normalization.  

\subsection{Binary angular momenta}
\label{sec:angmom}
Strong encounters (resonances and prompt exchanges) essentially randomize the angular momentum of a given binary, making a Fokker-Planck approach ill-equipped to deal with this aspect of the problem.  However, in the time between strong encounters, the cumulative effect of many weak flybys {\it will} lead to diffusive evolution of an individual binary's angular momentum, an effect largely neglected in existing approaches to binary evolution in dense clusters.  We therefore treat the problem of binary angular momentum in a two-timescale form.  An individual binary will undergo a random walk in angular momentum space {\it until} it suffers a strong binary-single scattering, at which point its new angular momentum can be drawn from the relevant distribution.  We will thus arrive at a steady state angular momentum distribution by solving a 1D Fokker-Planck equation in $e_{\rm B}$-space.  This {\it population-level} approach uses the diffusion coefficients computed in \citet[][their Eqs. 25a/25b]{hamers19b} to account for weak, distant encounters, and the mildly superthermal outcome distribution of \citet{stoneleigh19}, also found by \citet{ginat22}:
\begin{equation}
    \mathcal{N}_{\rm res}(e_{\rm B}) = \frac{{\rm d}\sigma}{{\rm d}e_{\rm B}} = \frac{6}{5}e_{\rm B}(1+e_{\rm B}) \label{eq:superthermal}
\end{equation}
as the ``initial conditions'' in a source term $S^+(e_{\rm B}) = \Gamma \mathcal{N}_{\rm res}(e_{\rm B})$ that accounts for the generation of new binaries after resonant encounters (a sink term $S^-(e_{\rm B}) = -\Gamma \mathcal{N}(e_{\rm B})$ is likewise used to remove existing binaries in resonant encounters).  Here we take a gravitationally focused strong scattering rate 
\begin{equation}
    \Gamma = \frac{2\pi G m_{\rm tot} n a_{\rm B}}{\sigma}, \label{eq:gammaEvaluated}
\end{equation}
i.e. the integral of Eq. \ref{eqn:gammanick} over an isotropic Maxwellian velocity distribution.
Note that in all cases we remain in the equal-mass limit, so that $m_{\rm tot}=3m$.  For practical calculations, it is somewhat easier to use the variable $\mathcal{R} = 1-e_{\rm B}^2$, which can be viewed as a dimensionless angular momentum.  Combining the resonant source/sink terms with the perturbative time evolution terms, we have
\begin{align}
    \frac{\partial \mathcal{N}}{\partial t} = & -\frac{\partial}{\partial \mathcal{R}}(\langle \Delta \mathcal{R} \rangle \mathcal{N}) + \frac{1}{2}\frac{\partial^2}{\partial^2 \mathcal{R}}(\langle (\Delta \mathcal{R})^2 \rangle \mathcal{N}) - S^-_e + S^+_e.  \label{eq:FP3}
\end{align}
Unlike Eq. \ref{eq:FP2}, the diffusion coefficients in this Fokker-Planck equation reflect the cumulative effect of many weak, perturbative flybys rather than that of repeated strong scatterings.  The diffusion coefficients $\langle\Delta \mathcal{R} \rangle$ and $\langle(\Delta \mathcal{R})^2 \rangle$ are derived in Appendix \ref{app:DCs} using the results of \citep{hamers19b} as a starting point.
We use a Dirichlet-type $\mathcal{N}=0$ boundary condition at high eccentricity $e_{\rm B} = e_{\rm coll}$, i.e. $\mathcal{R}=1-e_{\rm coll}^2$ (corresponding to collisions, or, in the case of black hole binaries, gravitational wave inspirals), and a zero-flux boundary condition at $e_{\rm B}=0$ ($\mathcal{R}=1$).  
By evolving this PDE forward in time, we can find a population-level steady state solution for arbitrary initial conditions.  

Alternatively, we can take initial conditions corresponding to the {\it outcomes} of the strong scatterings (Eq. \ref{eq:superthermal}), and evolve the eccentricity distribution forward under the influence of weak scatterings.  This {\it snapshot-level} approach describes the time evolution of an ensemble of binaries since their last strong scattering.  While the population-level approach provides astrophysically realistic eccentricity distributions, the snapshot-level approach is more useful for building physical understanding, specifically by disaggregating the effects of strong and weak scatterings.

\begin{figure}
\begin{center}
\includegraphics[width=\columnwidth]{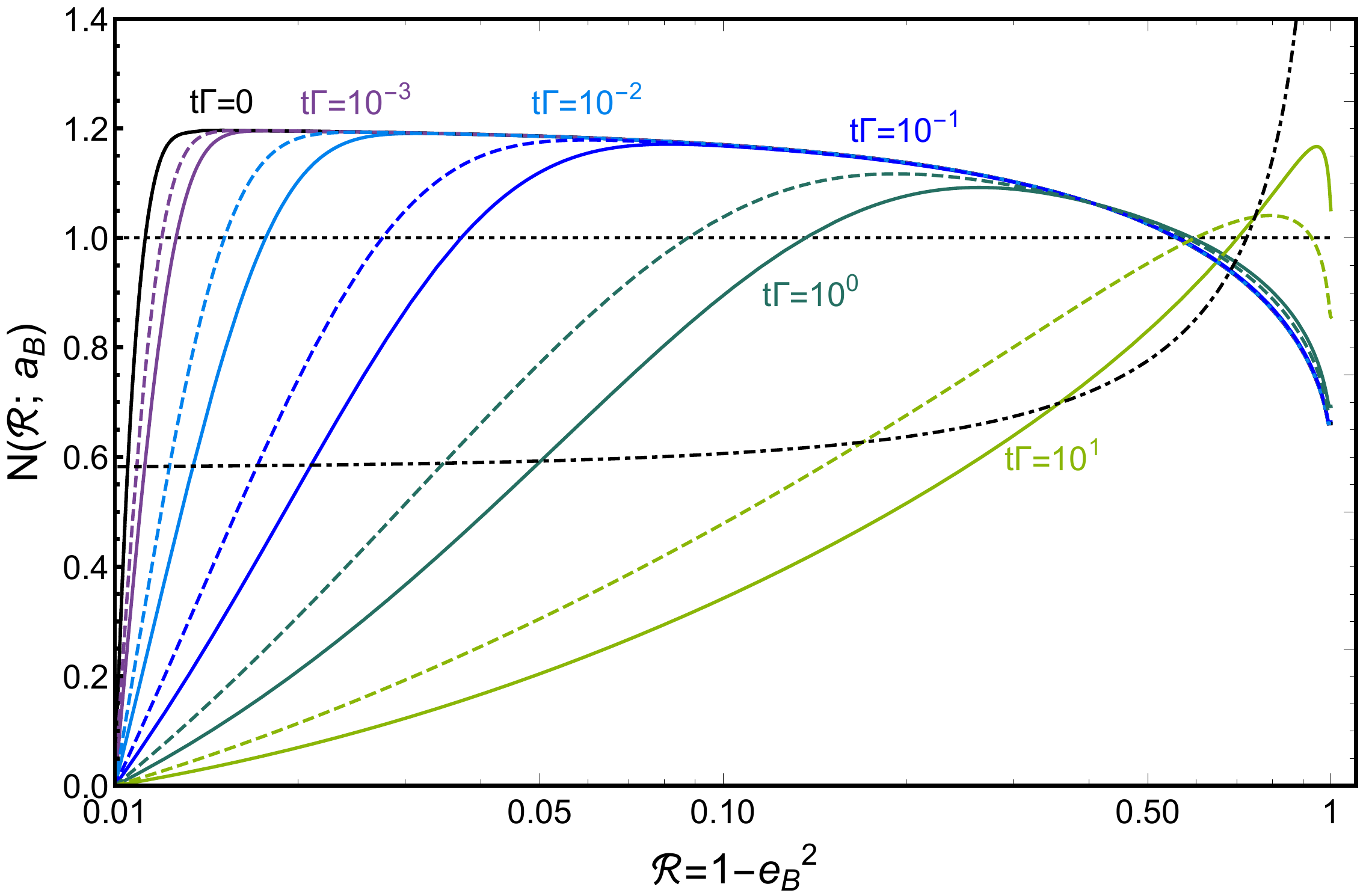}
\end{center}
\caption{Time evolution of the 1D Fokker-Planck equation in the space of dimensionless square angular momentum $\mathcal{R} = 1-e_{\rm B}^2$ ($\mathcal{R}=1$ is a circular orbit; $\mathcal{R}=0$ a radial one).  Here we show the number of stars per unit square angular momentum $\mathcal{N}(\mathcal{R}; a_{\rm B})$ at fixed semimajor axis $a_{\rm B}$.  We have assumed that the stellar radius $R_{\rm coll} = 10^{-2}a_{\rm B}$ (i.e. there is an absorbing boundary condition at $1-e_{\rm B}=10^{-2}$).  Different colors are labeled according to the fraction of the time between a strong (resonant) binary-single scattering event, which resets the binary eccentricity distribution to the mildly superthermal result discussed in the text.  However, the cumulative (diffusive) effect of many weak flybys will quickly evolve this superthermal input distribution into a more complicated one, with high-$e_{\rm B}$ (low-$\mathcal{R}$) orbits strongly depleted by direct collisions.  The subset of uncommon binaries that survive for more than a few resonant encounter times will achieve a distribution that is significantly more sub-thermal than the ${\rm d}N/{\rm d}e_{\rm B} \propto e_{\rm B}^{-4/25}$ distribution predicted in the absence of collisions \citep[][shown as a dot-dashed black line]{hamers19b}. For comparison, the usual thermal eccentricity distribution is shown as a dotted black line.  Our results depend modestly on $Q_{\rm min}$, the critical tertiary pericenter that is assumed to separate resonant from non-resonant encounters.  Solid colored lines show our fiducial value $Q_{\rm min}=3.4 a_{\rm B}$, while dashed colored lines show a more extreme choice of $Q_{\rm min}=2.4 a_{\rm B}$.  The net effect is modest.}
\label{fig:eccEvolution}
\end{figure}

Snapshot-level results are shown in Fig. \ref{fig:eccEvolution}.  We see that the initially super-thermal distribution (describing an ensemble of binaries shortly after their last resonant encounter) is initially eroded primarily by the Dirichlet boundary condition at high eccentricity.  After a time $t\sim \Gamma^{-1}$, however, the effect of many weak encounters serves to further redistribute binary orbits to the low-$e_{\rm B}$ side of the spectrum.  If we had considered weak scatterings only (i.e. zero-flux boundary conditions at both ends), we would have found a steady-state eccentricity distribution $\mathcal{N}(e_{\rm B}) \propto e_{\rm B}^{-4/25}$ \citep{hamers19b}, i.e. a moderately sub-thermal distribution biased towards circular orbits.  This analytic curve is shown for comparison in Fig. \ref{fig:eccEvolution}, but it is not achieved even for the rare subset of the binary population that survives for a time $10\Gamma^{-1}$ without a resonant encounter, attesting to the importance of the collisional boundary condition at high $e_{\rm B}$.  

\begin{figure}
\begin{center}
\includegraphics[width=\columnwidth]{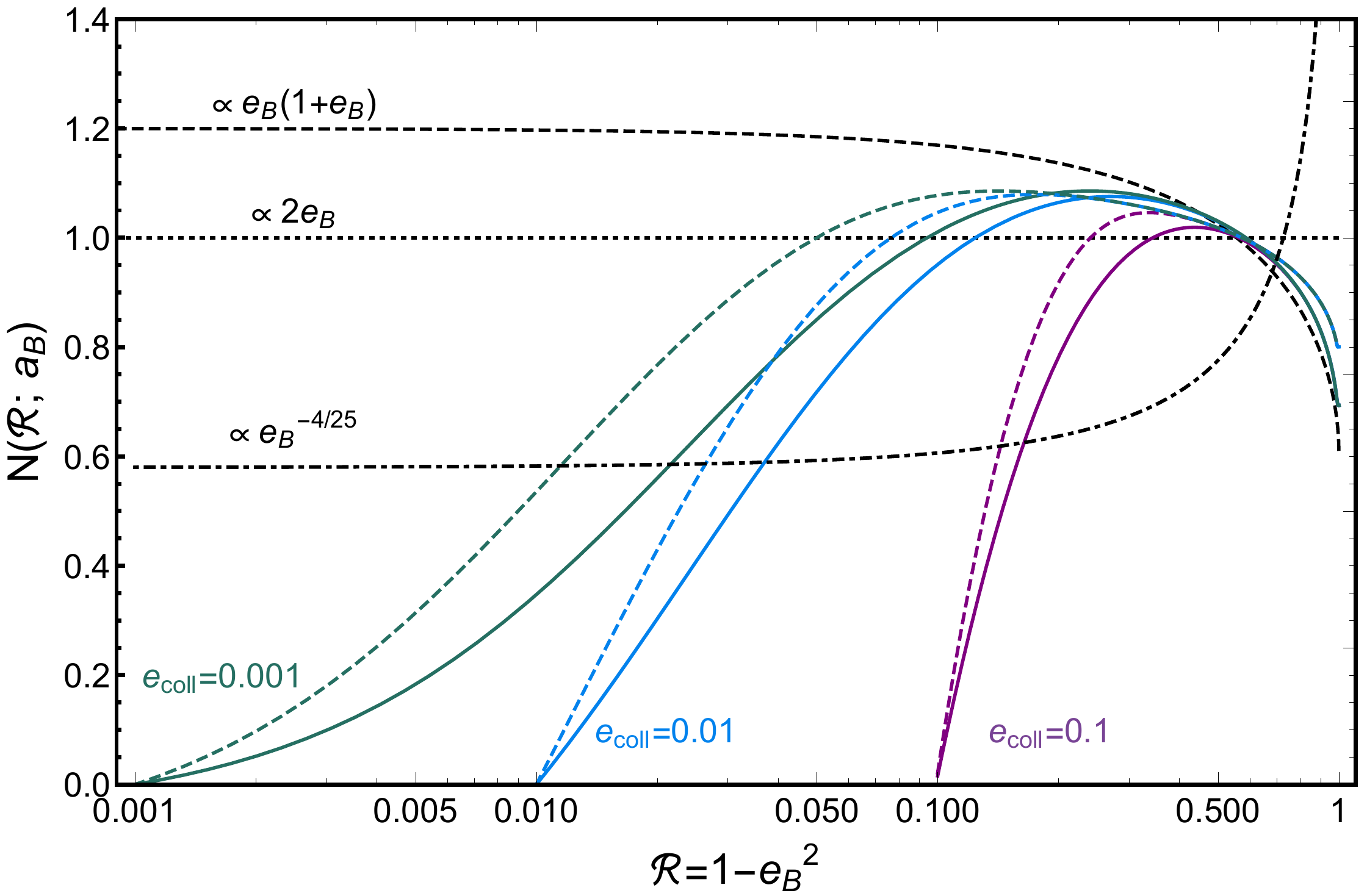}
\end{center}
\caption{Steady-state solutions of the 1D Fokker-Planck equation in the space of dimensionless square angular momentum $\mathcal{R}$.  Here we show the steady-state distributions achieved when combining (i) weak perturbations from secular flybys with (ii) less common resonant scatterings, whose outcomes are determined ergodically.  The solid green, blue and purple curves show steady-state distributions for populations of binaries that will merge when $\mathcal{R}$ equals $10^{-3}$, $10^{-2}$, and $10^{-1}$, respectively.  For comparison, the black dotted line is a thermal eccentricity distribution, the black dashed line is the superthermal ergodic outcome distribution for an ensemble of resonant scatterings, and the black dot-dashed line is the ${\rm d}N/{\rm d}e_{\rm B} \propto e_{\rm B}^{-4/25}$ distribution from \citet{hamers19b}.  In all cases the steady-state eccentricity (or $\mathcal{R}$) distribution is neither super- nor sub-thermal; instead it is peaked at intermediate eccentricity and depleted of both highly radial and highly circular orbits.  As in Fig. \ref{fig:eccEvolution}, solid colored lines represent $Q_{\rm min}=3.4 a_{\rm B}$, and dashed colored lines $Q_{\rm min}=2.4 a_{\rm B}$. }
\label{fig:eccDepletion}
\end{figure}

The combined effects of weak flybys, resonances, and a collisional boundary condition are fully visible in the population-level solutions in Fig. \ref{fig:eccDepletion}, which shows three different solutions for three different values of $e_{\rm coll}$.  Here we see that the steady-state $N(e_{\rm B})$ distributions are neither thermal, strictly super-thermal (as in \citealt{stoneleigh19}, nor strictly sub-thermal (as in \citealt{hamers19b}).  The distributions are peaked at an intermediate eccentricity $\sim 0.1-0.5$, but highly depleted at large $e_{\rm B}$ (due to the collisional boundary condition) and at nearly circular $e_{\rm B}$ (due to resonances).  We caution that our results do depend on the value of tertiary pericenter $Q_{\rm min}$ that transitions between resonant and non-resonant encounters.  Appendix \ref{app:DCs} contains a fuller discussion of this, but Figs. \ref{fig:eccEvolution} and \ref{fig:eccDepletion} demonstrate that this dependence is modest.

We also note here that a similar approach could be taken to the diffusive (weak scattering) and non-diffusive (resonant encounter) evolution of binary angular momentum orientations in a {\it non-isotropic cluster}.  Since resonant encounters preferentially produce survivor binaries with angular momentum vectors $L_{\rm B}$ aligned with the total angular momentum $\vec{L}_0$ of the three-body encounter, a non-isotropic velocity field will produce a preferential plane of binary rotation.  For example, a rotating star cluster with net angular momentum $\vec{L}_{\rm cl}$ would be expected to have binaries with orbital planes biased towards prograde alignment with $\vec{L}_{\rm cl}$.  In this first paper, however, we remain focused on the isotropic limit.  

\subsection{Energy backreaction to the host cluster} \label{backreaction}

In this section, we discuss and develop a toy model to quantify the effects of binary hardening via single-binary interactions for the energy content of the host star cluster.  Energy is systematically imparted to single stars via single-binary interactions, providing a heat source to the cluster core and subsequently further hardening those interacting hard binaries.  Integrating over the binary distribution, we can compute the flow of energy from this binary reservoir into the core, and its subsequent diffusion throughout the entire stellar system via two-body relaxation: indirect (predominantly long-range) single-single interactions.  

In principle, one could model the interactions between the binary population and the host cluster by adding extra dimensions to a Fokker-Planck approach.  The global evolution of star clusters is often treated in a Fokker-Planck way: assuming spherical symmetry, a population of stars (single- or multi-species) can be evolved forward in time in the space of orbital energy $\mathcal{E}$ and orbital angular momentum $\mathcal{L}$, with populations slowly diffusing due to two-body scatterings (indeed, solving this problem by probabilistically sampling stellar distributions is the basis of the Monte Carlo codes mentioned in \S \ref{sec:intro}).  While we hope to more rigorously explore this integrated approach in future work, for now we content ourselves to develop a ``two-zone'' model, where a star cluster with total mass $M_{\rm tot}$ is divided into a core of radius $r_{\rm c}$ and a halo extending further out, with a half-mass radius $r_{\rm h}$.  The cluster is collisionally relaxed and therefore is quasi-isothermal with a velocity dispersion $\sigma$ that is a weak function of radius.  For the sake of concreteness, we will use the analytic potential-density pair of \citet{stone15}.  In this three-parameter model ($M_{\rm tot}$, $r_{\rm h}$, $r_{\rm c}$), the mass density profile of the cluster is
\begin{equation}
    \rho(r) = \frac{\rho_{\rm c}}{(1+r^2/r_{\rm c}^2) (1+r^2/r_{\rm h}^2)},
\end{equation}
with a central core density of
\begin{equation}
    \rho_{\rm c} = \frac{M_{\rm tot}(r_{\rm c} + r_{\rm h})}{2\pi^2 r_{\rm c}^2 r_{\rm h}^2}
\end{equation}
and a core velocity dispersion 
\begin{equation}
    \sigma_{\rm c}^2 \approx \frac{6}{\pi} (\pi^2/8 -1) \frac{G M_{\rm tot}}{r_{\rm h}}.
\end{equation}
The approximate equality in the last expression indicates that this formula for the core velocity dispersion is taken in the $r_{\rm c} \ll r_{\rm h}$ limit (though it is correct to within $\sim 10\%$ for $r_{\rm c} \sim r_{\rm h}$).

We use this simple model to estimate the rate of orbital energy flow into the core (from binary-single scatterings) and out of the core (from two-body relaxation, primarily between single stars): $\dot{E}_{\rm c}={\rm d}E_{\rm c}/{\rm d}t$.  
The net flow is then
\begin{equation}
\label{eqn:energyflow}
\dot{E}_{\rm c} = \frac{{\rm d}E_{\rm c}}{{\rm d}r_{\rm c}}\frac{{\rm d}r_{\rm c}}{{\rm d}t} = \dot{E}_{\rm rel} + \dot{E}_{\rm bin}, 
\end{equation}
where $E_{\rm c} \approx$ -$M_{\rm c}\sigma_{\rm c}^2$/2 is the binding energy of the core, $r_{\rm c}$ is the core radius and $\sigma_{\rm c}$ is the 3D velocity dispersion in the core. Here the mass of the core is roughly (again, in the limit of $r_{\rm c} \ll r_{\rm h}$)
\begin{equation}
    M_{\rm c} \approx \frac{2(1-\pi /4)}{\pi} \frac{r_{\rm c}}{r_{\rm h}} M_{\rm tot}.
\end{equation}
The diffusive energy flow due to two-body relaxation is set by the radial gradient in ``temperature'' (i.e. velocity dispersion $\sigma$) across the cluster core and halo out to the half-mass radius.  In the limit of a single-mass cluster, it is roughly 
\begin{equation}
    \dot{E}_{\rm rel} = -A_{\rm cond} \frac{M_{\rm c}\sigma_{\rm c}^2}{2 t_{\rm r,c}},
\end{equation}
where $t_{\rm r, c}$ is the core relaxation time and $A_{\rm cond}$ is a dimensionless conductivity constant (i.e. an encapsulation of the very small gradient ${\rm d}\sigma / {\rm d}r$) that can be measured from Fokker-Planck (e.g. in single-mass clusters, $A_{\rm cond} \sim 10^{-3}$; \citealt{cohn80}) and N-body simulations.  In a multi-mass cluster, $A_{\rm cond}$ is higher by one to two orders of magnitude depending on the exact mass spectrum.  $\dot{E}_{\rm rel}$ is negative-definite because two-body relaxation conducts heat outwards, allowing the core to collapse and become more tightly bound. 

The rate of energy injection from binary burning can be computed by calculating the time evolution of the total energy in all binaries in the core,
\begin{equation}
    E_{\rm bin} = \int E_{\rm B} \mathcal{N}(E_{\rm B}) {\rm d}E_{\rm B}.
\end{equation}
As the binary burning rate is an integral over the distribution function, it turns Eq. \ref{eqn:energyflow} into an integro-differential equation.

One particularly interesting limit is when $\frac{dr_{\rm c}}{dt} \rightarrow 0$, since this defines the time of core "bounce", or the moment when core collapse halts and is reversed by single-binary interactions.  $N$-body and other approaches show that steady state solutions do not generally exist near this limit, which is the turning point in gravothermal oscillations.  

We now make a ``two-zone'' approximation for the binary populations, in which the binaries are divided into a portion in the cluster core (with distribution $\mathcal{N}_{\rm c}$) and a portion in the halo (with distribution $\mathcal{N}_{\rm h}$).  Note that since encounter rates scale as the local relaxation time, there are different diffusion coefficients for both the core (e.g. $\langle\Delta_{\rm c} E_{\rm B} \rangle$) and the halo (e.g. $\langle\Delta_{\rm h} E_{\rm B} \rangle$).  Binaries may move back and forth between these two zones via ejection from the core and dynamical friction on halo binaries.  These produce interchange rates that are 
\begin{equation}
    \frac{\partial N_{\rm ej}}{\partial t} = A_{\rm ej} \frac{1}{t_{\rm r,c}}
\end{equation}
and
\begin{equation}
    \frac{\partial N_{\rm sink}}{\partial t} = \frac{N_{\rm h}}{t_{\rm r,c}} \frac{r_{\rm c}^2}{r_{\rm h}^2}  \sqrt{\frac{m_{\rm B}}{\langle m_\star \rangle}},
\end{equation}
respectively.  In both of these equations, $t_{\rm r,c}$ is the core relaxation time (which increases by a factor $r_{\rm h}^2 / r_{\rm c}^2$ when one considers the relaxation time at the half-mass radius $r_{\rm h}$), and in the first, $A_{\rm ej}\sim 0.1-1$ is a dimensionless number computed by integrating over the Maxwellian velocity dispersion of the core stars.  

This gives a set of coupled differential equations for the time evolution of the cluster:
\begin{align}
    \dot{E}_{\rm c} =& \dot{E}_{\rm rel} + \dot{E}_{\rm bin} + \frac{1}{2}\sigma_{\rm c}^2\frac{\partial N_{\rm ej}}{\partial t} - \frac{1}{2}\sigma_{\rm c}^2\frac{\partial N_{\rm sink}}{\partial t}   \label{eq:twozone} \\
    \frac{\partial N_{\rm c}}{\partial t} =& -\frac{\partial}{\partial E_{\rm B}} \left(\langle \Delta_{\rm c} E_{\rm B} \rangle N_{\rm c} \right) + \frac{1}{2} \frac{\partial^2}{\partial E_{\rm B}^2} \left(\langle \Delta_{\rm c} E_{\rm B}^2 \rangle N_{\rm c} \right)\notag  \\
    +& \frac{\partial N_{\rm sink}}{\partial t} - \frac{\partial N_{\rm ej}}{\partial t} \notag \\
    \frac{\partial N_{\rm h}}{\partial t} =& -\frac{\partial}{\partial E_{\rm B}} \left(\langle \Delta_{\rm h} E_{\rm B} \rangle N_{\rm h} \right) + \frac{1}{2} \frac{\partial^2}{\partial E_{\rm B}^2} \left(\langle \Delta_{\rm h} E_{\rm B}^2 \rangle N_{\rm h} \right) \notag \\
    -& \frac{\partial N_{\rm sink}}{\partial t} + \frac{\partial N_{\rm ej}}{\partial t} \notag
\end{align}
Note that this is a system of two diffusive-type PDEs and one integro-differential ODE.  
\section{Results} \label{results}

In this section, we apply our model to dynamically evolve a population of binary star systems in a dense cluster environment and compare the results to N-body simulations.  We choose \"Opik's Law for our fiducial initial conditions, which gives the initial distribution of binary orbital energies.  We begin with a one-zone model, before presenting the results for our two-zone (core/halo) model, using Eqs. \ref{eq:twozone}.

\subsection{One-Zone Model}

We begin by dynamically evolving a population of binary star systems through energy space in a dense cluster environment, by numerically solving the Fokker-Planck equation derived in Section~\ref{master}.  
The initial distribution of binary orbital energies is again chosen (according to \"Opik's Law) to be $f_{\rm B}(E_{\rm B}) = k|E_{\rm B}|^{-1}$, where $k$ is a normalization constant.  The results have already been shown in Figure~\ref{fig:FokkerPlanckSolution} at several different core relaxation times $t_{\rm r,c}$, indicated by the different coloured lines, but here we will discuss them in more detail. 

In Figure~\ref{fig:FokkerPlanckSolution}, we see slow evolution of the hardest binaries, and relatively quick evolution of ones near the hard-soft boundary.  This is unsurprising, given the small (large) scattering cross-sections of the former (latter).  Over time, a quasi-steady state is reached at low binary energies, which slowly propagates to higher energies.  Once this quasi-steady state solution reaches the collisional (high-$|E_{\rm B}|$) boundary, the shape of the distribution freezes in, and further evolution only decreases its normalization.  At all times, the drift coefficient $\langle E_{\rm B} \rangle$ is creating a net flow towards softer energies, though the diffusion term $\langle (E_{\rm B})^2 \rangle$ is responsible for the turnover at the hardest energies, once a quasi-steady state has been established everywhere.  
\subsection{Two-Zone Model}

Next, we add many of the extra pieces of the two-zone model, i.e. Eq. \ref{eq:twozone}.  Because our goal is to eventually compare to $N$-body simulations with limited run-time (and limited evolution of the cluster density profile, i.e. limited global energy transfer), here we solve only the two coupled PDEs that exchange binary populations between the cluster core and halo.  Simultaneous solution of the energy equation would require more numerical development which we defer to future work.  
%

As we will show in the subsequent sections, we initially observe a quick depletion of binaries in the core, due to the recoil imparted by single-binary interactions, and a corresponding increase in the halo population.  At later times, these binaries mass segregate back into the core, increasing the population of binaries in the core and decreasing that in the halo.  Eventually, an approximate steady-state balance is achieved between the rate of binaries being ejected from the core due to single-binary interactions and binaries re-entering the core by leaving the halo due to mass segregation.  This is because we assume all equal-mass particles in our model, such that binaries are the heaviest objects in the cluster.

\subsection{Comparisons to N-body simulations} \label{comparisons}

In this section, we present the results of our preliminary N-body simulations, and compare them to the predictions of our second-order analytic models described in the previous sections, with a focus on the improved two-zone model.

\subsubsection{Initial Conditions and Assumptions}

For each simulation, we adopt a Plummer density profile initially, with 10$^4$ stars and 200 binaries.  The initial cluster has a core radius of 0.3 pc and a half-mass radius is 0.8 pc.  We assume identical point-particles with masses of 1 M$_{\odot}$.  As in our analytic calculations, we assume \"Opik's Law initially for the distribution of binary orbital energies and a thermal eccentricity distribution.  At the hard end, we truncate our initial energy distribution at a minimum value of 4 times the radius of the Sun (i.e., corresponding to slightly wider than a contact state for 1 M$_{\odot}$ stars).  At the soft end, we truncate at twice the hard-soft boundary, calculated as:
\begin{equation}
\label{eqn:HS}
a_{\rm HS} = \frac{Gm}{\sigma^2},
\end{equation}
where $\sigma$ is the core velocity dispersion, which is initially 2.0 km s$^{-1}$. 
The initial core and half-mass relaxation times for our simulated clusters are 7.6 Myr and 25 Myr, respectively. 

We perform 10 simulations all with the same initial conditions each perturbed slightly using a different random seed.  These simulations are then stacked together, to increase our sample size for the number of binaries evolving dynamically due to single-binary interactions, bringing the total sample size up to 2000.  This stacking is done to increase the statistical significance of our results, and verify the robustness of our N-body simulations by quantifying the stochastic contribution of chaos to the observed differences in each simulation.  

\begin{figure}
\begin{center}
\includegraphics[width=\columnwidth]{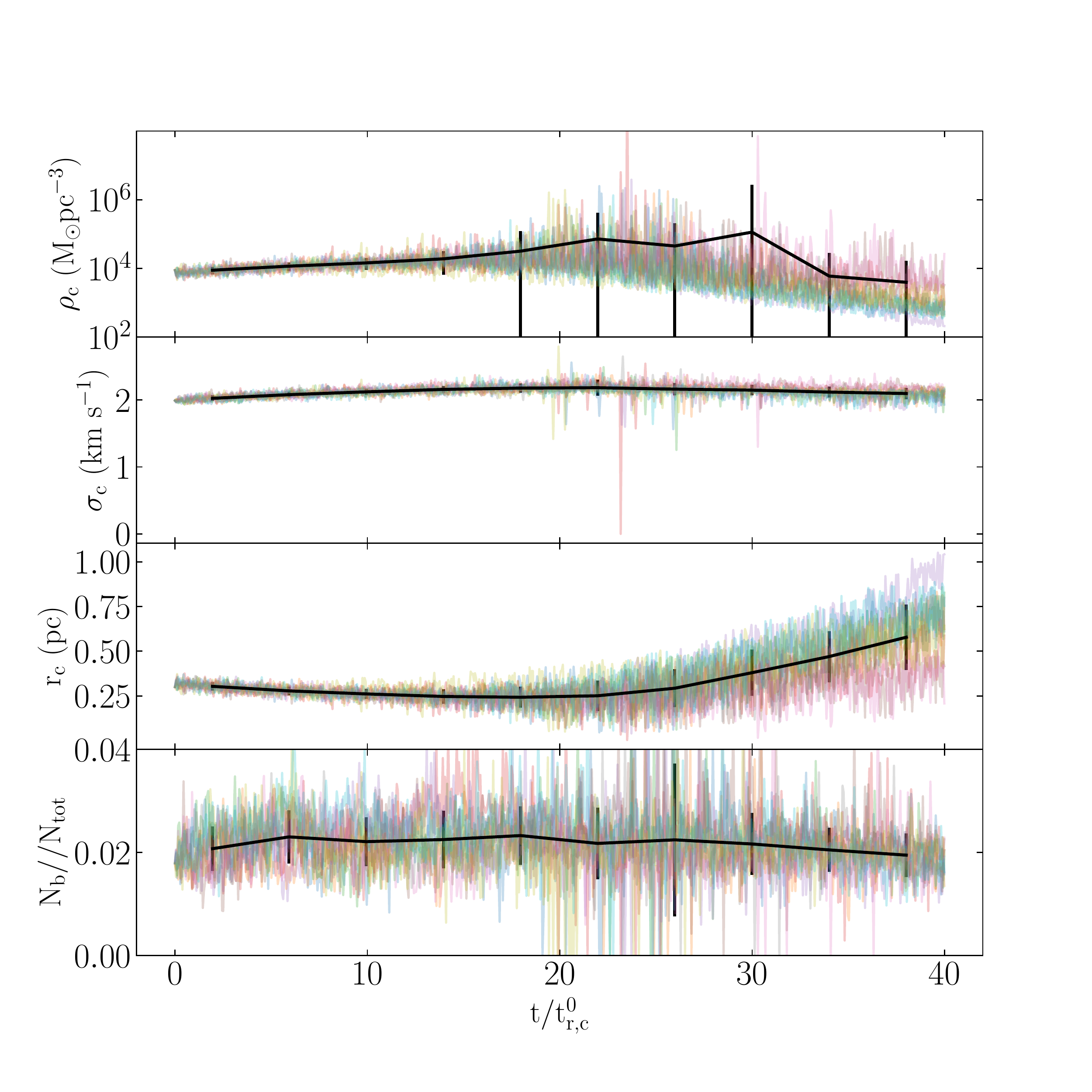}
\end{center}
\caption{The time evolution, normalized by the initial core relaxation time, of the core density, core velocity dispersion, core radius, and core binary fraction for all 10 simulations with the mean value illustrated as a solid black line. The error bars mark the standard deviation about the mean.} 
\label{fig:binplot}
\end{figure}

The time evolution of several core properties in each of the 10 simulations, namely the core density, velocity dispersion, radius, and binary fraction are illustrated in Figure \ref{fig:binplot}, with the average values illustrated in black. Time is normalized by the cluster's initial core relaxation time $t_{r,c}^0$ and the error bars represent the standard deviation about the mean. For $t/t_{r,c}^0 < 20$, the core evolution of all 10 models is very similar, with core density and velocity dispersion staying nearly constant while the core radius and binary fraction slowly decrease with time. However, near $t/t_{r,c}^0 = 20$, the clusters undergo a mild core collapse. In the post-core collapse stage, the core density and radius of individual simulations slowly start to diverge with significant fluctuations between timesteps. The core velocity dispersion, however, remains close to its original value with a few brief fluctuations and the core binary fraction slowly decreases beyond $t/t_{r,c}^0 = 20$ due to binary destruction, after most binaries outside of the core have had enough time to mass segregate into the core.  We emphasize that the comparison is most reliable before core-collapse occurs, due to the increased stochasticity in the time evolution of the cluster properties beyond this point.  In other words, the simulations begin to diverge significantly beyond core collapse, yielding increasingly different cluster properties between the simulations over time.  Also, our neglect of energy backreaction from the binary population becomes worse at this time.

\subsubsection{One-Zone Model}

\begin{figure}
\begin{center}
\includegraphics[width=\columnwidth]{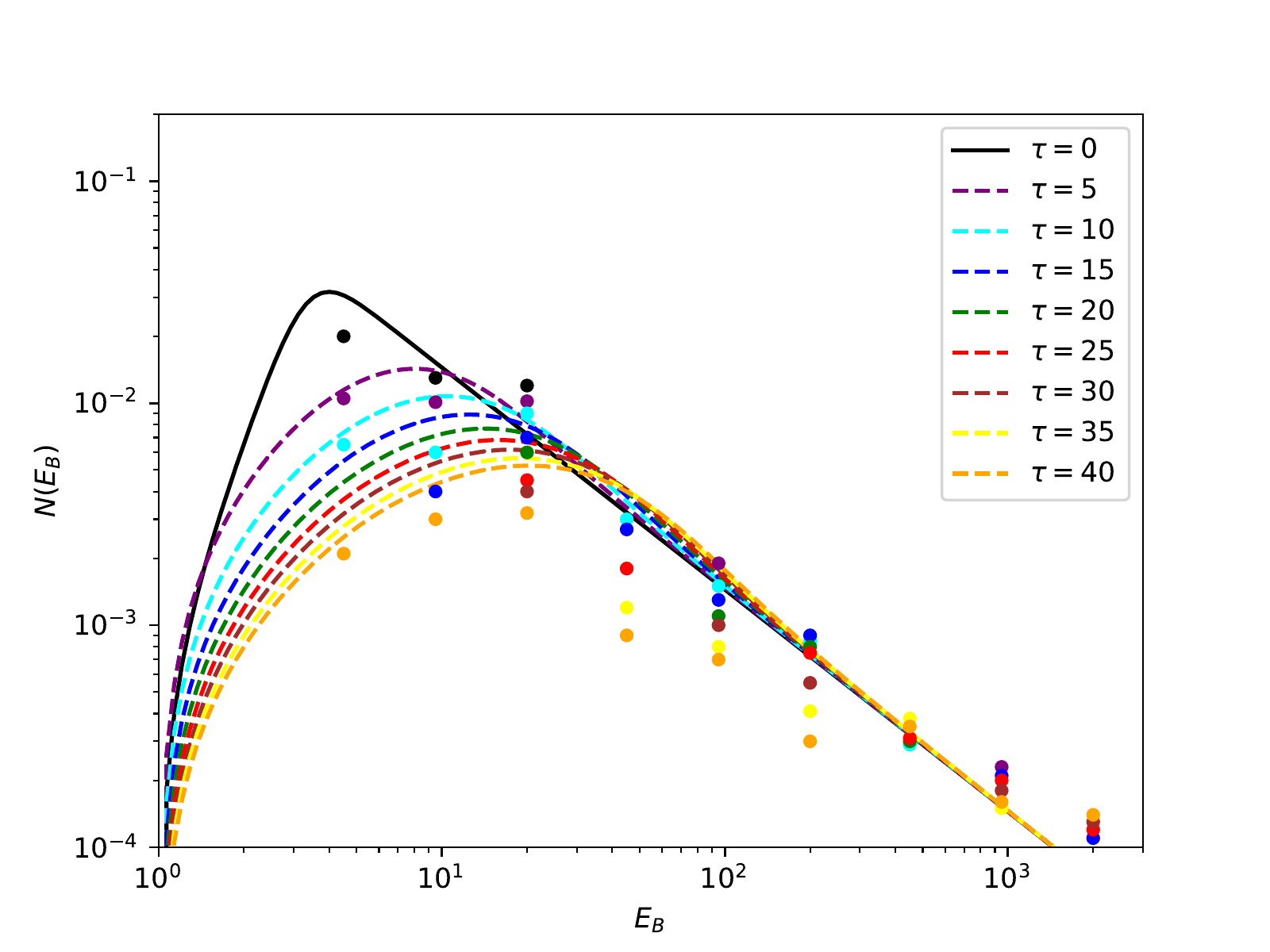}
\end{center}
\caption{The distribution of binary energies in the core of the cluster, shown at nine different snapshots in time.  Black indicates the initial distribution corresponding to $t = 0t_{\rm r,c}^0$, where $t_{\rm r,c}^0$ is the initial core relaxation time.  We then evolve our simulation for in total 45 core relaxation times, in steps of 5 $t_{\rm r,c}^0$. We plot the distributions for each time-step using different colours, as indicated in the upper-right inset in the figure. Data points are binned-up binary energies from the N-body simulations, while curves are predictions from the one-zone Fokker-Planck approach.} 
\label{fig:onezone}
\end{figure}

In this section, we discuss the time evolution of the binary orbital energy distribution for our one-zone model, focusing on the cluster core.  The binary orbital energy distributions of our N-body simulations are shown in Figure~\ref{fig:onezone} at several different core relaxation times.  The black solid line show the initial distribution given by \:Opik's Law.

The time evolution of the binary orbital energy distribution behaves as expected, slowly pushing hard binaries to become more compact and soft binaries to become disrupted.  The evolution of the one-zone model is initially faster than the simulations as $N(E_B)$ quickly decreases after just 5 core relaxation times. It is not until 40 core relaxation times does $N(E_B)$ for the simulations fall below the one-zone model. This discrepancy is likely because binaries flow into the core from the halo due to two-body relaxation in our simulations, and out of the core due to the recoil imparted post-single-binary interaction.  But, in our analytic model, we assume that anything happening in the core stays in the core.  Hence, binaries are depleted in the core at a faster rate, both due to mergers at the hard end of the distribution and the destruction of wide binaries at the soft end of the distribution.

After a given number of core relaxation times, Figure~\ref{fig:onezone} illustrates a strong agreement between the one-zone model and the simulations at the hard- and soft-ends of the distribution. However, the model over-predicts the number of binaries in the intermediate regime relative to the simulations by a factor $\sim$ 3 (when error bars are included this typically means that the N-body simulation results disagree with our analytic model at the level of $\gtrsim$ 3$\sigma$).  We attribute this disagreement to the overall faster core evolution in our one-zone model, depleting binaries in the intermediate regime by pushing hard binaries to become harder, and wide binaries to become wider.  In the subsequent section, we will move on to our two-zone model, and show that this does indeed correct this disagreement.

\subsubsection{Two-Zone Model}

In this section, we discuss the time evolution of the binary orbital energy distribution for our two-zone model, considering now both the core and the halo.  Binaries are now allowed to flow out of the core in our analytic model due to the recoil imparted by linear momentum conservation post-single-binary interaction, and later flow back into the core due to mass segregation.

\begin{figure}
\begin{center}
\includegraphics[width=\columnwidth]{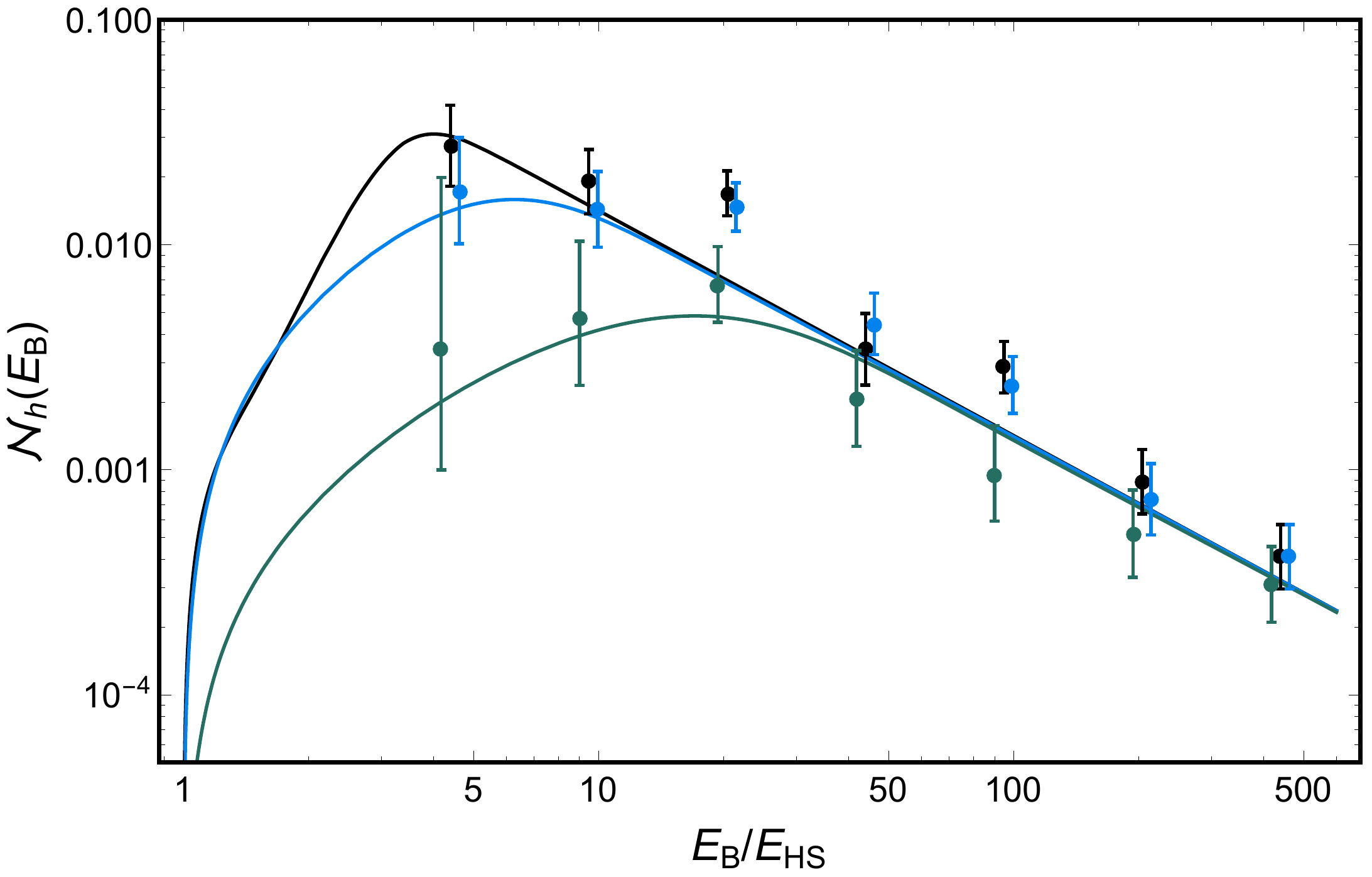}
\end{center}
\caption{The distribution of binary energies in the halo of the cluster, shown at three different snapshots in time.  Black, blue, and green correspond to $t=0 t_{\rm r,c}^0$, $t=10 t_{\rm r,c}^0$, and $t=40 t_{\rm r,c}^0$, respectively.  Data points are binned-up binary energies from the N-body simulations (error bars show asymmetric $1\sigma$ Poissonian error range; \citealt{Gehrels86}), while curves are predictions from the two-zone Fokker-Planck approach.} 
\label{fig:binCompareHalo}
\end{figure}

\begin{figure}
\begin{center}
\includegraphics[width=\columnwidth]{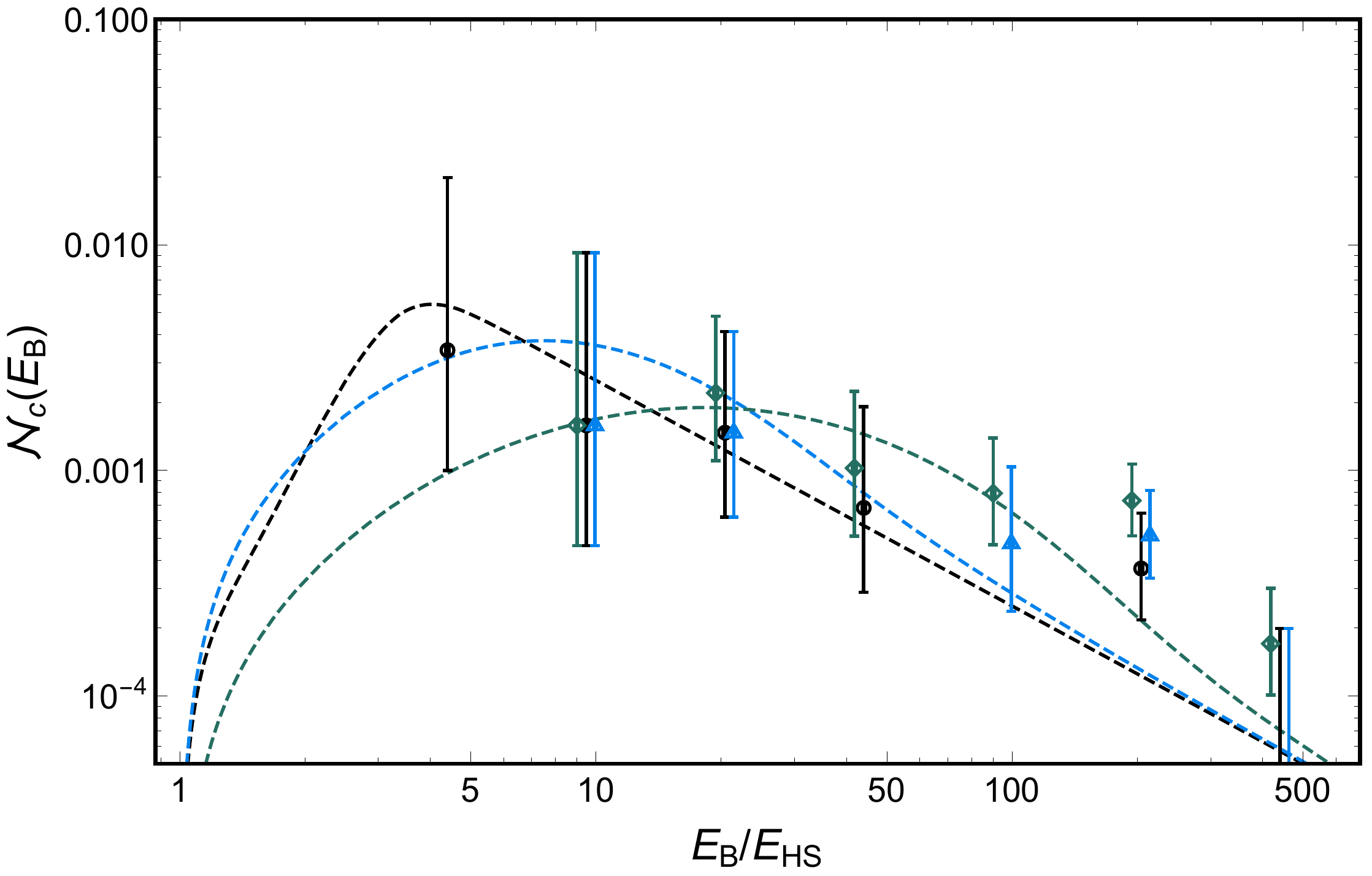}
\end{center}
\caption{The same as Figure~\ref{fig:binCompareHalo}, but for the cluster core, where statistics are poorer.} 
\label{fig:binCompareCore}
\end{figure}

As is clear from Figures~\ref{fig:binCompareHalo} and~\ref{fig:binCompareCore}, the results from our N-body simulations, showed by the coloured data points, now agree with our analytic two-zone model typically to within 1$\sigma$, with only a few outliers within 2$\sigma$ from our analytic model.  We conclude that our improved two-zone model does indeed improve the agreement between the simulations and our analytic theory.  We emphasize that this is but one of the many changes that can be accommodated by our model, to even further improve the agreement between the simulated and calculated results.  Possible improvements and how to implement them will be discussed in more detail in the subsequent section.

\section{Discussion} \label{discussion}

In this section, we discuss the key features and assumptions going into our model, their short-comings and future efforts that should be made for constructing an improved analytic model for dynamically evolving forward through time populations of binary star systems in dense stellar environments due to single-binary interactions.  We further discuss the significance of our analytic models and their results to pertinent problems characteristic of modern astrophysics.

\subsection{A new tool independent of computational methods for dynamically evolving populations of binaries} \label{newtool}

The Boltzmann-type equation derived in this paper presents an alternative and complementary tool to computational $N$-body or Monte Carlo simulations.  These methods compute the time evolution of their binaries using a combination of stellar dynamics, stellar and binary evolution, etc.  The method presented here isolates the dynamical effects of single-binary scatterings on the time evolution of binary orbital parameter distribution functions.  Thus, our method is meant to be directly complementary to analogous computer simulations.  In principle, comparing the results of the model presented here with more complicated $N$-body and Monte Carlo simulations should allow us to better isolate the effects of dynamics in evolving a binary population in a given environment, thereby isolating the effects of stellar and binary evolution, etc. (i.e., usually included in $N$-body and MC models in an approximate way, assuming the evolution proceeds in isolation), while also comparing the relevant timescales for each of these critical effects to operate.  Thus, our model allows us to quantify and characterize the dominant physics evolving a given binary population in a given star cluster at any time, by rigorously quantifying and isolating the effects of single-binary interactions.

Recently, \citet{geller19} took a significant step forward in this direction by developing a semi-analytic Monte Carlo-based model for the dynamical evolution of binaries that is analogous to a first-order diffusion-based model.  
They compared the results to direct $N$-body and Monte Carlo models for GC evolution with very similar initial conditions, and found excellent agreement over several Gyr of cluster evolution.  This model was subsequently applied in \citet{trani21} to quantify the impact of binary-binary scatterings on the results, since the authors considered only single-binary interactions.  As illustrated in this paper, adopting the higher-order diffusion-based model presented herein only serves to improve the agreement between the analytic theory and computational simulations.

\subsection{Improving the model} \label{future}

Next, we discuss the key features and assumptions going into our model, and how they can be improved upon in future work.

First, we begin with a one zone model for our calculations, including only those binaries and singles inside the cluster core.  However, in reality, objects are free to migrate in and out of the core.  In particular, we expect the core binary fraction to initially increase over time due to binary stars mass segregating into the core due to two-body relaxation.  We also expect that the probability of a given binary being ejected from the core due to the recoil from linear momentum conservation during a single-binary interaction will be higher for more compact binaries.  These are likely to eject singles at higher velocities, imparting a larger recoil velocity to the binary and increasing the probability of ejecting it from the core, or even the cluster.  

All of these aspects of our model can easily be improved upon by adopting more realistic boundary conditions.  One way to go about this is to increase the number of zones, and expanding the model to include regions of the cluster outside of the core (e.g., adopting a shell-like structure for the cluster).  As illustrated in Figures~\ref{fig:binCompareHalo} and~\ref{fig:binCompareCore}, we have shown that adapting our model to a two-zone model does indeed improve the agreement, treating the core and halo independently.  This allows for binaries to flow into and out of the cluster core and halo, due to being ejected from the core via single-binary interactions and subsequent mass segregation back into the core from the halo.  

We also do not allow for binaries to be kicked out of the cluster, since in our simulations these events are very rare.  However, for those clusters where such events are more likely, additional sink/source terms could be included in our existing Boltzmann equation, improving the accuracy of our existing two-zone model even further.

Second, we do not consider binary formation via three-body interactions involving initially all isolated single stars, since these events are exceedingly rare in our simulations (or do not occur at all).  This effect should become important primarily at very high densities, such as during core collapse or in very massive, dense star clusters and galactic nuclei.
In future work, this aspect of our model can be improved upon by including an additional density-dependent source term in our Boltzmann equation.  

Third, we have assumed that the properties of the core are constant (i.e., a constant density and velocity dispersion throughout the core at a given time) in our analytic model since this is approximately true for the initial cluster conditions considered in this paper.  However, we manually update the averaged core properties at regular intervals upon performing our comparisons to the $N$-body simulations.  The inclusion in our analytic model of a more realistic gradient in the core gravitational potential would introduce a dependence of the encounter rate on the distance from the cluster centre $r$.  In future work, this aspect of our model can be improved upon by including a radial $r$ dependence into Equation~\ref{eqn:gammanick}.  We do not expect this to have significantly affected the comparison between our analytic model and the results of our $N$-body simulations, since we terminate the comparison before the point at which most simulations reach core collapse and begin to diverge in their time evolution.

Fourth, we have neglected binary-binary interactions, which should also always be occurring for non-zero binary fractions, and will hence contribute to the dynamical evolution of the binary orbital properties.  This should not have significantly affected the results of our comparison between our analytic model and the simulations.  This is because single-binary interactions occur more frequently than binary-binary interactions for binary fractions $\lesssim$ 10\% \citep{leigh11}, and we intentionally adopt very low initial binary fractions in our simulations to minimize the contribution of binary-binary interactions relative to single-binary interactions.  We compensate by performing additional simulations in order to generate the required statistical significance (i.e., the total number of simulated binaries) for the comparisons between our analytic model and the simulations, by stacking the results of these simulations all having nearly identical initial conditions.  This aspect of our work can be improved upon in future studies, using the methodology described in the Appendix.

Finally, we assume identical mass particles throughout this paper.  However, our model can easily be improved to include a mass-dependence using, for example, the method described in Section 3.1 of \citet{leigh20}.

\subsection{Astrophysical implications:  A focus on black hole binaries in globular clusters} \label{BHs}

It turns out that the assumptions needed for the methods and modeling techniques presented in this paper to be valid are particularly well-suited to treat the dynamics of stellar-mass black hole (BH) binaries in globular clusters.  This is because the assumptions underlying the application of our analytic formalism for single-binary scatterings require low virial ratios - i.e., the total kinetic energy must be a small fraction of the total binding energy of the interaction, to help maximize the fraction of long-lived, chaotic, resonant interactions, for which our formalism is most directly applicable (i.e., the assumption of ergodicity is upheld).  Due to energy equipartition, BH binaries can end up with the lowest velocities in the cores of clusters, and also tend to have very large absolute orbital energies due to the larger component masses.  Both effects push us toward preferentially low virial ratio interactions, ideal for applying the methods introduced in this paper to study the dynamical evolution of BH-BH binaries in dense star clusters \citep{leigh16b}.  We intend to explore this potentially interesting connection and application for our model in forthcoming work.


Finally, it is also in principle possible to use our model to constrain the primordial properties of star clusters.  In particular, if a given cluster is observed to be in core-collapse and the observer can measure the binary population properties (i.e., the present-day number of binaries and their orbital parameter distributions), then one can constrain the initial conditions using our model.  This is because our model assumes a given set of initial conditions, and these become a good candidate for the true set of initial conditions if the model reaches core-collapse on the correct timescale.  Degeneracies in the initial conditions yielding approximately the same time of core collapse are likely to exist, but these can easily be identified and quantified using our analytic model in future work, which indicates one of its main strengths and utility for astrophysical research: a quick and fast exploration of the relevant parameter space to help constrain the underlying sets of initial cluster conditions that could have evolved over a Hubble time to reproduce what we observe today (i.e., central density, total mass, surface brightness profile, etc.).

\section{Summary} \label{summary}

Moore's Law is dead.  
It follows that the demand for alternative models independent of computational limitations is increasing, and the field of gravitational dynamics is no exception. With this in mind, we recently derived, using the ergodic hypothesis \citep{monaghan76a,monaghan76b}, analytic outcome distributions for the products of single-binary \citep{stoneleigh19} scatterings (similar techniques may also apply to binary-binary scatterings, \citealt{leigh16b}).  With these outcome distributions in hand, it becomes possible to construct a diffusion-based approach to dynamically evolve an entire population of binaries due to single-binary (and eventually including binary-binary) scatterings.

We present in this paper a self-consistent statistical mechanics-based analytic model formulated in terms of a master equation (in the spirit of \citealt{goodman93}), and eventually evolved in its Fokker-Planck limit.  Our model evolves the binary orbital parameter distributions in dense stellar environments forward through time due to 
strong single-binary interactions, using the analytic outcome distribution functions found in \citet{stoneleigh19}.  The effects of weaker, perturbative scatterings are incorporated using the secular theory of \citet{hamers19a, hamers19b}.

We have applied our formalism in various simplified limits, working for now in the equal-mass case.  In the space of binary eccentricity $e_{\rm B}$, we find that the combined effect of strong (resonant) scatterings and more distant, perturbative flybys is to create steady-state $e_{\rm B}$ distributions that are strongly depleted at both the lowest and highest eccentricities.  The resulting binary populations do not match the strongly subthermal $e_{\rm B}$ distributions arising from weak scatterings only \citep{hamers19b}, nor the mildly superthermal distributions coming from strong scatterings only \citep{stoneleigh19}, but rather are peaked at an intermediate eccentricity $e_{\rm B} \sim 0.1-0.5$.

In the space of binary energy, we compare the predictions of our semi-analytic model to the results of numerical $N$-body simulations performed using the NBODY6 code.  We find good agreement between the simulations and our analytic model for the initial conditions considered here, and the adopted time intervals over which the results are compared (i.e., the first 20 core relaxation times, and roughly up until the time at which core collapse occurs).    

The semi-analytic model presented in this paper represents a first step toward the development of more sophisticated models, which will be the focus of future work.  For example, as shown in Section~\ref{backreaction}, it is in principle possible to 
couple binary evolution equations to a few-zone model to produce physically transparent models of cluster core collapse and gravothermal evolution.  More speculatively, the binary evolution formalism presented here could be applied to study the collisional evolution of binaries in (simplified models of) dense star systems with fewer degrees of symmetry, where $N$-body simulations can be prohibitively expensive and Monte Carlo techniques do not currently work.  Rotating star clusters (though see \citealt{fiestas06}) and inclination-segregated stellar disks \citep{meiron19} are two examples of such systems. 

In future work, we intend to further expand upon the base model presented in this paper to include binary-binary interactions.  These can occur more frequently than single-binary interactions in clusters with high binary fractions (i.e., binary-binary interactions dominate over single-binary interactions for binary fractions $\gtrsim$ 10\% \citep{sigurdsson93,leigh11}), and hence cannot be neglected in this cluster regime (e.g., open clusters and low-mass globular clusters).  In an Appendix, we consider this added complication to our base model, touching upon some basic predictions that motivate the need for further development in this direction.  Specifically, we identify a potential steady-state balance between the binary and triple fractions in star clusters, based on simple thermodynamics-based considerations, and show via a proof-of-concept calculation that this could in principle be used to directly constrain the initial primordial binary and triple fractions, in addition to the primordial properties of the multiple star populations in dense stellar environments.  

\section{Acknowledgments}

NWCL gratefully acknowledges the generous support of a Fondecyt Iniciaci\'on grant 11180005, as well as support from Millenium Nucleus NCN19-058 (TITANs) and funding via the BASAL Centro de Excelencia en Astrofisica y Tecnologias Afines (CATA) grant PFB-06/2007.  NWCL also thanks support from ANID BASAL project ACE210002 and ANID BASAL projects ACE210002 and FB210003.  NCS received financial support from the Israel Science Foundation (Individual Research grant 2565/19), and the BSF portion of a NSF-BSF joint research grant (NSF grant No. AST-2009255 / BSF grant No. 2019772). WL acknowledges support from NASA via grant 20-TCAN20-001 and NSF via grant AST-2007422.  

\appendix
\section{Adapting to the four-body problem and binary-binary scatterings} \label{higherorder}

Due to their non-negligible binary fractions ($\gtrsim$ 1\%), the dynamical modification of the binary populations in open and globular clusters must necessarily include not only single-binary interactions but also direct binary-binary interactions, and possibly even interactions involving triples (see \citet{leigh13} and \citet{leigh14b} for more details).  The contribution to this evolution from four-body interactions increases with increasing binary fractions.

This can be understood by drawing an analogy between stellar multiplicity in star clusters and the formation/destruction of molecules in an isothermal gas (see \citet{leigh13} and \citet{geller15} for more details).  
In a hot gas, collisions between particles are energetic and occur frequently, contributing to the destruction of molecules and the formation of more atoms and ions.  This is in direct analogy with the destruction of triples and binaries in dynamically hot star clusters (i.e., having high velocity dispersions), and a stark reduction in the fractions of stellar multiples.  In a cold gas, on the other hand, particle-particle collisions tend to be of low energy, allowing for particles to more easily "stick together".  This stimulates the production of molecules, and increases the relative fractions of molecules and atoms/ions in the gas.  This is in direct analogy with the preservation of binaries and even the formation of higher-order stellar multiples (e.g., triples produced during binary-binary interactions) in dynamically cold star clusters (i.e., having low velocity dispersions), having high multiplicity fractions.  

In this section, we explore and discuss how the model presented in this paper for single-binary scatterings can be adapted to also treat binary-binary scatterings.

\subsection{Accommodating an additional particle} \label{nis4}

In this section, we briefly consider how the model presented in this paper could be adapted to include binary-binary interactions (more generally, we consider temporarily gravitationally-bound configurations of 3 \textit{and} 4 particles).  That is, we consider the role of binary-binary interactions in dynamically evolving an initial distribution of binary orbital energies forward through time. We save the detailed derivations for future work, and instead sketch out rough toy models meant to highlight the potential of further pursuing this line of research.  As we will show, we predict a steady-state balance between the fractions of binaries and triples in dense star clusters (in the limit that the host cluster properties are static in time).  

As already described, any analytic modeling seeking to accurately quantify the time evolution of a population of binary stars in a dense star cluster must include not only single-binary interactions, but also binary-binary interactions.  Both types of interactions occur in such clusters, both affecting the orbital parameter distributions.  Above a critical binary fraction (i.e., the fraction of unresolved point sources that are binaries) of $\sim$ 10\%, binary-binary encounters dominate over single-binary encounters, with the latter occurring at a higher rate \citep{sigurdsson93,leigh11}.  Given that open clusters are much more numerous than globular clusters in the Milky Way, binary-binary interactions dominate over single-binary interactions in all environments but the most massive, densest star clusters (with the lowest binary fractions due to their high velocity dispersions, and hence small orbital separations corresponding to the hard-soft boundary which contribute to enhancing soft binary disruption).

The first step in extending the model presented in this paper for single-binary scatterings to binary-binary scatterings is the introduction of additional ``macroscopic'' outcomes (i.e., comprising different numbers of bound and/or unbound triplets, pairs and singles).  This is because single-binary scatterings involving point-particles and negative total interaction energies only ever end by producing a final ejected single star and a binary star system, unlike the analogous binary-binary scatterings. The latter have three possible outcomes, if all four objects are point particles:  (1) a binary and two ejected single stars; (2) a stable hierarchical triple and an ejected single star; and (3) two binary stars.  The introduction of these additional macroscopic states demands the development of a more sophisticated Boltzmann-type equation (i.e., treatment of the underlying dynamical evolution) than is necessary for single-binary scatterings, since each outcome must be treated individually.  In the subsequent section, we will discuss a number of key features that should result from an improved model and, ultimately, the vast increase in available phase space (given the addition of a fourth particle).

\subsection{A steady-state dynamical balance between binaries and triples} \label{steady}

In this section, we focus on an illustrative example centered on the formation/destruction of dynamically-stable triple systems during binary-binary scatterings, in a given host cluster trying to achieve a ``steady-state'' balance (i.e., binaries and triples are created/destroyed at the same rate).  

\begin{figure}
\begin{center}
\includegraphics[width=\columnwidth]{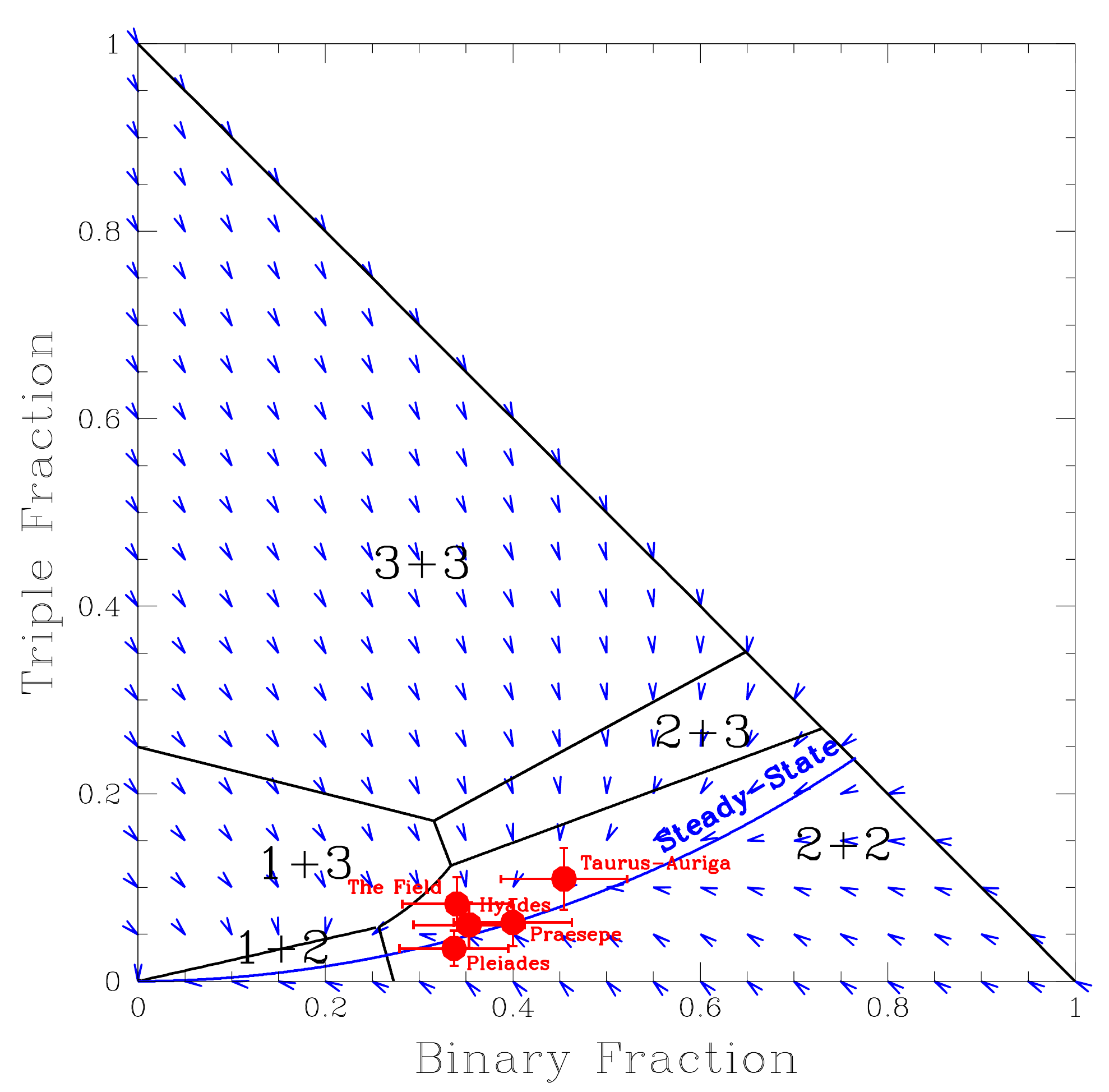}
\end{center}
\caption[A Multiplicity Interaction Rate Diagram (MIRD) showing the locations of the steady-state binary and triple fractions]{Illustrative plot showing the cluster triple fraction as a function of the binary fraction, or a Multiplicity Interaction Rate Diagram. The solid circles show the observed data for several open clusters and the Galactic Field, taken from \citet{leigh13}.  
The solid blue lines shows the steady-state solution for our choice of input parameters, as described in the text. The blue arrows show the flow line directions in the binary fraction-triple fraction-plane, calculated from our suite of numerical scattering experiments.}
\label{fig:fig2}
\end{figure}

In Figure ~\ref{fig:fig2} we show the Multiplicity Interaction Rate Diagram (MIRD) used throughout the following analysis. The MIRD shows the parameter space in the binary fraction-triple fraction-plane for which each type of dynamical interaction (i.e., single-single or 1+1, single-binary or 1+2, single-triple or 1+3, etc.) dominates. 
That is, the observed binary and triple fractions can be measured for a particular star cluster, and the resulting data point can be placed on the MIRD. Whatever region of the plot the data point falls on immediately indicates the type of dynamical interaction that currently dominates over all others in that cluster.  As is clear from Figure ~\ref{fig:fig2}, for low-mass star clusters in the Milky Way, the rates of interactions involving singles, binaries and triples all occur at comparable rates (see \citet{leigh11} for more details), but binary-binary interactions dominate in all clusters (and the field) considered here.

The MIRD shown in Figure~\ref{fig:fig2} is constructed as follows. First, the solid black lines segment the parameter space in the $f_{\rm b}-f_{\rm r}$-plane for which each type of dynamical interaction dominates. Briefly, these lines are calculated by equating the analytic encounter rate estimates from \citet{leigh11} for each type of interaction, and solving to obtain a relation between the binary and triple fractions (see \citet{leigh11}, \citet{leigh13} and \citet{leigh18} for more details). We assume $R = 1 R_{\odot}$, $a_{\rm b} =$ 3 AU and $a_{\rm t} =$ 10 AU for the mean single star radius, binary orbital separation and outer-most triple orbital separation, respectively. Although somewhat arbitrary, these values are thought to be representative of typical binaries and triples in real open clusters (OCs) and GCs \citep[e.g.][]{leigh13,geller15}. We note that the MIRD shown in Figure~\ref{fig:fig2} is meant only as an illustrative example. More realistic assumptions can of course be adopted as our Boltzmann formalism is further developed to include binary-binary scatterings, allowing us to perform more detailed and robust calculations of the steady-state relation between binary and triple fractions and for application of specific MIRDs tailored to individual clusters.

Second, the blue arrows show the direction of flow in the binary fraction-triple fraction-plane corresponding to the time evolution of these two parameters, for a given initial combination. These arrows are calculated using a combination of analytic interaction rates and numerical scattering simulations, and correspond to the instantaneous rates (i.e. over an infinitesimally small amount of time). For the indicated initial binary and triple fractions, the number of each type of interaction is calculated over some fixed interval of time (taken to be much smaller than the cluster age).

The numbers of single, binary and triple stars produced in these interactions are then calculated using numerical simulations of 1+2, 2+2, 1+3, 2+3 and 3+3 interactions performed with the \texttt{FEWBODY} code (see \citet{fregeau04} and \citet{leigh12} for more details about the code). The contribution from each type of interaction is then scaled according to the corresponding analytic rate. For these simulations, we assume identical point particles with masses of 1 M$_{\odot}$. For simplicity, all binaries have separations of 1 AU initially.  The initial inner and outer orbits of all triples have separations of, respectively, 1 AU and 10 AU. The relative velocity at infinity is sampled uniformly between 0 and 0.5$v_{\rm crit}$, where $v_{\rm crit}$ is the critical velocity, defined as the relative velocity at infinity corresponding to a total encounter energy of zero. The initial impact parameter in our calculations is always set to zero, which yields sufficiently informative results for our illustrative purposes here  \citep[e.g.][]{leigh16c,geller19}.  To calculate these instantaneous vectors in $f_{\rm b}-f_{\rm t}$-space, we obtain the branching ratios (i.e., the fraction of our simulations producing singles, binaries and triples, for any type of interaction, specifically 1+2, 2+2, 2+3, and so on) directly from our numerical scattering experiments.  For example, the approximate fractions of singles, binaries and triples produced from 2+2 scatterings can be seen in Figure 4 of \citet{leigh16b}, and yield for low virial ratios (i.e., corresponding to our chosen initial conditions for the scattering simulations performed here) $f_{\rm t}$ $\sim$ $f_{\rm b}$ $\sim$ 10\% and $f_{\rm s} \sim$ 80\%.

Finally, the solid blue line shows, for a given binary fraction, the predicted triple fraction when the host cluster is in steady-state. That is, when the rates of triple creation and destruction via dynamical interactions are in approximate balance. To calculate the steady-state line, we once again use a combination of our analytic interaction rates and numerical scattering simulations, as described above. We assume that triples can only be created during 2+2 interactions \citep{littlewood52}, and can only be destroyed during 1+3, 2+3 and 3+3 interactions.  \textit{Interestingly, all OCs in the sample from \citet{leigh13} are approximately 1-$\sigma$ away from the steady-state line, suggesting that the binary and triple fractions in these clusters are technically consistent with being in approximate thermal equilibrium according to our model assumptions.}  More work is needed to better understand the underlying physics, and how the observed features of the MIRD is informing us about the underlying time evolution (since each MIRD represents a static snapshot of the host cluster evolution in this parameter space).  Among other things, a MIRD should be made for each cluster individually using its own observed cluster properties.  This work will be addressed in a forthcoming paper.

If we assume that \textit{all} triples are destroyed during every 1+3, 2+3 and 3+3 interaction, then the criterion for steady-state for the binary and triple fractions can be written:
\begin{equation}
\label{eqn:btsteady}
f_{\rm 2+2}\Gamma_{\rm 2+2} = f_{\rm 1+3}\Gamma_{\rm 1+3} + f_{\rm 2+3}\Gamma_{\rm 2+3} + f_{\rm 3+3}\Gamma_{\rm 3+3},
\end{equation}
where f$_{\rm 2+2}$, f$_{\rm 1+3}$, f$_{\rm 2+3}$ and f$_{\rm 3+3}$ are the fractions of, respectively, 2+2, 1+3, 2+3 and 3+3 encounters that produce stable hierarchical triples. 
In spite of the simplifying assumptions that went into making Figure~\ref{fig:fig2}, it reveals a potentially rich and interesting dynamical evolution that deserves further study.  As already discussed, a more accurate and robust way forward in understanding the full scale of this richness could be to construct a Boltzmann equation including \textit{both} the 1+2 and 2+2 cases.  With this, it should be possible to more formally (and hopefully more accurately) calculate a predicted steady-state line in the binary fraction-triple fraction-plane.  To achieve this, the required machinery for 1+2 interactions is now entirely in-hand and can be found in \citet{stoneleigh19}.  For 2+2 interactions, for which the parameter space of possible outcomes is much larger, the procedure begun in \citet{leigh16b} can be used as a rough guide moving forward.  Subsequent papers already in progress on the four-body problem will deliver the required analytic solutions for each of the three possible outcomes of the chaotic four-body problem.  With these in hand, our Boltzmann equation can be completed and used for parameter space explorations, which will be the focus of future work.

\subsection{Understanding the physics of a ``steady-state'' multiplicity solution, and the potential for using it to constrain the initial conditions} \label{communal}

Many previous studies trying to understand the dynamical evolution of populations of binaries relied on the assumption of steady-state. That is, that the time-derivative of the global distribution function is equal to zero \citep[e.g.][]{heggie75}.  But do the multiple star populations in most real clusters ever reach steady-state within a Hubble Time, roughly independent of the initial conditions? If they do, on what characteristic time-scale does this occur?  This is an important question since, if all clusters should have reached steady-state by the present-day independent of their initial conditions, then this could make it very difficult to trace back their previous dynamical evolution and constrain their primordial birth environments.  Conversely, if no clusters should have reached steady-state by the present-day, then both Figure~\ref{fig:fig1} and Figure~\ref{fig:fig2} illustrate that the observed present-day multiplicity properties can be used to directly constrain the corresponding initial conditions.  This is because the flow lines in the binary fraction-triple fraction-plane do not overlap.  Hence, a given region in this parameter space can only be accessed by specific initial combinations of f$_{\rm b}$ and f$_{\rm t}$.  In other words, not every region in this parameter space is causally connected via the underlying dynamical evolution, making it in principle possible to use the presently observed binary and triple fractions in a given star cluster to directly constrain the primordial birth properties of the host cluster.

\section{Angular momentum diffusion}
\label{app:DCs}
Here we describe how we use the formalism of \citet{hamers19b} to compute the secular diffusion coefficients $\langle \Delta \mathcal{R} \rangle $ and $\langle (\Delta \mathcal{R})^2 \rangle $ that are employed in \S \ref{sec:angmom}.

Our starting point here are the diffusion coefficients $\langle \Delta \tilde{\mathcal{R}} \rangle$ and $\langle (\Delta \tilde{\mathcal{R}})^2 \rangle$ taken from Eqs. 25a/25b in \citet{hamers19b}.  These diffusion coefficients are functions of the dimensionless binary angular momentum $\mathcal{R}$ and also of the small ``secular approximation parameter''
\begin{equation}
    \epsilon_{\rm SA} = \frac{1}{4\sqrt{3}} \left(\frac{a_{\rm B}}{Q} \right)^{3/2},
\end{equation}
where $a_{\rm B}$ is the semimajor axis of the binary, $Q$ is the pericenter of the distant perturber with respect to the binary center of mass, and we have taken (i) equal-mass encounters and (ii) parabolic perturber orbits.  Assumption (i) is restrictive and should be relaxed in future work, but assumption (ii) is a reasonable approximation in the limit of hard binaries.  We have denoted the \citet{hamers19b} diffusion coefficients with over-tildes because they depend explicitly on perturber pericenter $Q$, unlike the strong-scattering diffusion coefficients used elsewhere in this paper, which have been implicitly or explicitly averaged over a range of tertiary pericenters.  Our task here is to perform the same procedure for these secular diffusion coefficients.  

Integrating over a range of tertiary impact parameters $b$, we can calculate a tertiary-averaged diffusion coefficient
\begin{align}
    \langle (\Delta \mathcal{R})^n \rangle =& \int_{\rm b_{\rm min}}^\infty \pi b \langle (\Delta \tilde{\mathcal{R}})^n \rangle {\rm d}b \\
    \approx& \frac{2\pi n G m_{\rm tot} a}{\sigma} \int^\infty_{Q_{\rm min}} \langle (\Delta \tilde{\mathcal{R}})^n \rangle{\rm d}Q \notag \\
    =& \Gamma \int^\infty_{q_{\rm min}} \langle (\Delta \tilde{\mathcal{R}})^n \rangle{\rm d}q. \notag
\end{align}
The approximate equality in the second line represents the gravitationally focused limit (relevant for the hard binaries we are focused on), and in the third line we simplify the dimensional prefactor by changing variables from $Q\to q = Q/a$ and identifying the strong scattering rate $\Gamma$ (Eq. \ref{eq:gammaEvaluated}).  Unfortunately, these integrals are generally dominated by contributions from tertiaries with small $Q$, necessitating a cutoff at some minimum pericenter $Q_{\rm min}$ (or equivalently $b_{\rm min}$).  Physically, it is tempting to identify this with the transition between resonant and non-resonant encounters, which in the equal-mass limit occurs at a value\footnote{While there is of course not an immediate transition between resonances and flybys based solely on tertiary pericenter $Q$ (other initial conditions contribute as well), the transition visible in the end-state maps of \citet{samsingilan18} is surprisingly abrupt, so we take this numerical result as a reasonable approximation.} of $q \approx 3.4$ \citep{samsingilan18}.  We thus take $q_{\rm min}=3.4$ as a fiducial value, but examine an alternate case in \S \ref{sec:angmom} and find generally mild effects for plausible ranges of $q_{\rm min}$.

With this averaging formalism specified, we may now integrate $\langle \Delta \tilde{\mathcal{R}} \rangle$ and $\langle (\Delta \tilde{\mathcal{R}})^2 \rangle$ over tertiary pericenter.  In contrast to the results of \citet{hamers19b}, we convert rational numbers to decimal notation for brevity.  In the end, we find:
\begin{align}
    \langle \Delta \mathcal{R} \rangle  =& -(1-\mathcal{R}) \bigg(q_{\rm min}^{-2} \Big(1.866\mathcal{R}-0.7931 \Big) \\
    &+ q_{\rm min}^{-5}\Big( 0.5586\mathcal{R}^2 - 0.5640\mathcal{R}+0.1699\Big) \bigg) \notag
\end{align}
and
\begin{align}
    \langle (\Delta \mathcal{R})^2 \rangle  =& (1-\mathcal{R})^2 \bigg(q_{\rm min}^{-2} \Big(1.542\mathcal{R}\Big) \\
    &+ q_{\rm min}^{-5}\Big( 6.485\mathcal{R}^2 - 4.294\mathcal{R}+0.6795\Big)\notag \\
   &+ q_{\rm min}^{-8}\Big(5.725\mathcal{R}^3 - 7.212\mathcal{R}^2 + 3.246\mathcal{R} - 0.5152\Big)\notag \\
   &+ q_{\rm min}^{-11}\Big(1.146\mathcal{R}^4 - 2.303 \mathcal{R}^3 + 1.902\mathcal{R}^2 - 0.7485 \mathcal{R} \notag \\
   &+0.1190\Big)\notag\bigg) \notag.
\end{align}
These diffusion coefficients reflect secular torques from passing stars up to second order in the parameter $\epsilon_{\rm SA}$.

\end{document}